\documentclass[twocolumn,aps,pre,superscriptaddress,floatfix]{revtex4-1}
\usepackage{graphicx}
\usepackage{epsfig}
\usepackage{thumbpdf}
\usepackage{amsfonts}
\usepackage{times}
\usepackage{float}
\usepackage{indentfirst} 
\usepackage{amsmath}
\usepackage{epstopdf}
\usepackage{textcomp}
\usepackage{tocvsec2}
\usepackage{epsfig}
\usepackage{xcolor}
\usepackage{comment}
\usepackage{CJK} 

\usepackage[section]{placeins}
\makeatletter\AtBeginDocument{%
     \expandafter\renewcommand\expandafter\subsection\expandafter
       {\expandafter\@fb@subsecFB\subsection}%
     \newcommand\@fb@subsecFB{\FloatBarrier
     \gdef\@fb@afterHHook{\@fb@topbarrier \gdef\@fb@afterHHook{}}}
     \g@addto@macro\@afterheading{\@fb@afterHHook}
     \gdef\@fb@afterHHook{}
  }
\usepackage[colorlinks=true,linkcolor=black]{hyperref} 

\newcommand{\be}{\begin{equation}}
\newcommand{\ee}{\end{equation}}

\begin{document}
\begin{CJK}{UTF8}{gbsn}

\title{Close and ordinary social contacts: how important are they in promoting large-scale contagion?}

\author{Peng-Bi Cui (崔鹏碧)}
\affiliation{School of Computer Science and Engineering, University of Electronic Science and Technology of China, Chengdu 610054, China}
\affiliation{Institute of Fundamental and Frontier Science, University of Electronic Science and Technology of China, Chengdu 610073, China}
\affiliation{Big Data Research Center, University of Electronic Science and Technology of China, Chengdu 611731, China}

\author{Wei Wang}
\affiliation{Cybersecurity Research Institute, Sichuan University, Chengdu 610065, China}

\author{Shi-Min Cai}\email{shimin.cai81@gmail.com}
\affiliation{School of Computer Science and Engineering, University of Electronic Science and Technology of China, Chengdu
610054, China}
\affiliation{Institute of Fundamental and Frontier Science, University of Electronic Science and Technology of China, Chengdu 610073, China}
\affiliation{Big Data Research Center, University of Electronic Science and Technology of China, Chengdu 611731, China}

\author{Tao Zhou}
\affiliation{School of Computer Science and Engineering, University of Electronic Science and Technology of China, Chengdu
610054, China}
\affiliation{Big Data Research Center, University of Electronic Science and Technology of China, Chengdu 611731, China}
\affiliation{Institute of Fundamental and Frontier Science, University of Electronic Science and Technology of China, Chengdu 610073, China}

\author{Ying-Cheng Lai}
\affiliation{School of Electrical, Computer, and Energy Engineering, Arizona State University, Tempe, Arizona 85287-5706, USA}

\begin{abstract}

An outstanding problem of interdisciplinary interest is to understand 
quantitatively the role of social contacts in contagion dynamics. In general, 
there are two types of contacts: close ones among friends, colleagues and 
family members, etc., and ordinary contacts from encounters with strangers. 
Typically, social reinforcement occurs for close contacts. Taking into account 
both types of contacts, we develop a contact-based model for social contagion. 
We find that, associated with the spreading dynamics, for random networks 
there is coexistence of continuous and discontinuous phase transitions, but 
for heterogeneous networks the transition is continuous. We also find that 
ordinary contacts play a crucial role in promoting large scale spreading, and  
the number of close contacts determines not only the nature of the phase 
transitions but also the value of the outbreak threshold in random networks.
For heterogeneous networks from the real world, the abundance of close 
contacts affects the epidemic threshold, while its role in facilitating the 
spreading depends on the adoption threshold assigned to it.
We uncover two striking phenomena. First, a strong interplay between ordinary 
and close contacts is necessary for generating prevalent spreading. In fact,
only when there are propagation paths of reasonable length which involve both 
close and ordinary contacts are large scale outbreaks of social contagions 
possible. 	
Second, abundant close contacts in heterogeneous networks promote both outbreak 
and spreading of the contagion through the transmission channels among the 
hubs, when both values of the threshold and transmission rate among ordinary 
contacts are small.
We develop a theoretical framework to obtain an analytic understanding of the 
main findings on random networks, with support from extensive numerical 
computations. Overall, ordinary contacts facilitate spreading over the entire 
network, while close contacts determine the way by which outbreaks occur, i.e.,
through a second or first order phase transition. These results provide 
quantitative insights into how certain social behaviors can emerge and become 
prevalent, which has potential implications not only to social science, but
also to economics and political science.

\end{abstract}

\date{\today}
\maketitle
\end{CJK}

\section{Introduction} \label{sec:intro}

The spreading of certain behaviors (contagion) in the human society 
has social, economical, and political implications, which 
has attracted a great deal of interdisciplinary research effort. Conventional
methods developed to uncover and understand the dynamics of social contagion
are more or less based on some kind of threshold and memory
effects~\cite{Watts2002,Dodds2004,Centola2007,Yaugan2012,Weiss2014,Nematzadeh2014,Lee2014,Karsai2014,Centola2018,Guilbeault2018}. 
For example, it is likely for an individual to adopt certain behavior 
if he/she possesses friends who have already adopted the behavior. From a
network point of view, for a node to adopt certain behavior, the number of
connected nodes who have already accepted and exhibit the behavior must 
exceed a threshold~\cite{Watts2002,Ruan2015}. That is, an individual will
become willing to adopt a behavior if he/she has received sufficient and
repeated information about the behavior from his/her friends (or neighbors 
in the underlying social network)~\cite{Cui2014,Wang2015}. 
In the real world, the influences of network neighbors and the tendency for
any individual to adopt certain behavior can be highly non-uniform. There 
is empirical evidence that individuals and their social contacts tend to play 
heterogeneous roles in contagion~\cite{Barrat2007,Karsai2016}.
For example, regardless of the nature of the behavior, there always exist 
certain individuals who are reluctant to accept or adopt the behavior. To 
account for the heterogeneity, a variant of the classic threshold 
model~\cite{Watts2002} was introduced~\cite{Ruan2015}, where a certain fraction 
of nodes were assumed to be immune to the behavior. These are the blocked 
nodes, and if their number or density is large enough, both the extent and
diffusion speed of the contagion spreading will be suppressed, provided that 
no cascading or avalanche type of processes occur. The effects of  
heterogeneity in social contacts have been recently studied~\cite{Zhu2017}
using a contagion threshold model incorporating weighted edges, with the 
finding that heterogeneity in the weights can suppress the propagation of 
contagion.

The influences from connected neighbors in a social network represent a kind 
of reinforcement effect, where the probability of adopting certain social 
behavior by an individual is modified when he/she receives information about 
the behavior through social contacts. In recent years, spreading dynamics 
driven by reinforcement have attracted a great deal of attention from 
researchers in diverse fields including social science, economics, and 
physics~\cite{Centola2010,Centola2018,Lu2011,Piedrahita2013,Hodas2014,Cui2014}.
{\em The key fact that motivated our present work is that, with respect to
reinforcement, the nature of social contacts can have drastically
different impacts}. In particular, from close social contacts such as family
members, friends, and colleagues, reinforcement can be much stronger 
than that from conventional or ordinary social contacts with, e.g., 
strangers. For a model of social contagion to capture the real
behaviors as accurately as possible, the distinct reinforcement effects 
from close and ordinary social contacts must be taken into account.    

The need to distinguish two types of social contacts in terms of their
influences has been well documented in the literature. Historically, 
Lazarsfeld and Merton pointed out that mass media messages 
can be reinforced by interpersonal communications~\cite{Lazarsfeld1948}. 
There were empirical evidence and theories for the conjecture that more 
interpersonal conversation or discussions can promote the impact of media 
information through reinforcement~\cite{Chaffee1988,Scheufele2002}. 
In general, individuals are more likely to engage in interpersonal 
communications with family, friends or colleagues as often as they get media 
news or messages from newspapers or other individuals~\cite{Desmet2015}. 
That is, for reinforcement, close social contacts are more 
effective than interaction with machines or strangers. At the present, there 
exists considerable empirical evidence that, even for interpersonal 
communications, there is heterogeneity in their role in social reinforcement. 
In general, the degree of reinforcement depends on factors such as individual 
responsiveness, the number of neighbors capable of reinforcing, and 
respondent-reinforcer pairing~\cite{Vollmer2001,Pierce2013,Skinner2015}. 
For example, the number of reinforcers is determined by the degree of the 
individual in the social network, and reinforcement among strangers, even if 
there are social contacts among them, is far less likely than that among close 
relationships~\cite{Skinner2015}. That is, in reality, social contagion is 
strongly contact-based, and not all social contacts can lead to reinforcement. 

To our knowledge, in the current literature, there is no work on social 
contagion dynamics which takes into account the two distinct types of social
contacts: close and ordinary. In general, both types of contacts exist, and 
the question is how their coexistence affects the basic dynamical behaviors. 
In this paper, we propose a general model of social contagion with two 
distinct transmission channels: one through close and another through ordinary 
contacts. For random networks, we develop an analytical edge-based 
compartmental method to solve the model, which enables us to make a number 
of predictions in terms of the fundamental characterizing quantities such as 
the final outbreak size, the outbreak threshold, and the nature of the phase 
transition.We also carry out extensive agent-based, stochastic simulations 
to assess the performance of the model. For random networks, the computations 
reveal that continuous (second order) 
and discontinuous (first order) phase transitions coexist in random networks, 
which is ascertained by an analytic bifurcation analysis of the system.
A computational study of two representative empirical networks with a 
heterogeneous degree distribution from the real world reveals that only 
continuous transitions can be expected.
We provide a physical understanding of the basic spreading dynamics through a 
detailed statistical analysis, uncover the conditions under which contagion 
prevalence can arise, and validate the existence of an optimal fraction of 
ordinary contacts for outbreak at a global scale. 

Two striking phenomena are uncovered. One is that an interplay between 
ordinary and close contacts is necessary for generating prevalent spreading 
on random networks. In particular, only when there are propagation paths of 
sufficient length which involve both close and ordinary contacts are large 
scale outbreaks of social contagions possible. 
The second phenomenon is that, for heterogeneous networks, a considerable 
number of close contacts will promote outbreak of the contagion by forming 
channels for successful transmissions among hubs. Taken together, ordinary 
contacts make possible spreading over the entire network, while close 
contacts not only shape the way by which outbreaks occur, i.e., through 
a second or first order phase transition in random networks, but also 
facilitate local spreading and outbreaks in heterogeneous networks. 
These results provide quantitative insights into how social behaviors can 
emerge and become prevalent, which has potential implications not only to 
social science, but also to economics and political science.

In Sec.~\ref{sec:model}, we present our general model of spreading dynamics
with two distinct types of social contacts. In Sec.~\ref{sec:theory}, we 
describe the edge-based compartmental analysis approach. In 
Sec.~\ref{sec:results}, we implement our spreading model on random networks 
and carry out agent-based simulations and a comprehensive theoretical analysis, 
and then extend our study to empirical heterogeneous networks. 
In Sec.~\ref{sec:discussion}, we present conclusions and an outlook.

\section{Model} \label{sec:model}

To gain basic insights into the roles of close and ordinary contacts in social
contagion dynamics, we firstly assume that the individual social relationships 
are described by a random network of size $N$ (i.e. Erd\"{o}s-R\'{e}nyi (ER) 
networks), where a pair of nodes are connected with each other with 
probability $p_{e}$ so that the network degree distribution and the mean 
degree are $p(k)=e^{-\langle k\rangle}\langle k\rangle^k/k!$ and 
$\langle k\rangle=Np_e$, respectively. The total number of edges is
$E=N\langle k\rangle/2$, where each edge represents a particular social 
contact between two individuals. The dynamics of social contagion are governed
by the standard susceptible-adopted-recovered (SAR) model, where any
node (individual) can be in one of the three states. In particular, a node 
in the susceptible state may adopt a behavior when it receives information 
about it from its neighbors who have already adopted the behavior. If a 
node is in the adopted state, with certain probability it will transmit 
the information to its susceptible neighbors. A node in the recovered state 
is ``idle'' and does not transmit any information. A unique feature, which 
makes our model distinct from the classic SAR model, is that there are two 
distinct ways for an adopted node to transit information to a susceptible 
neighbor, depending on whether the underlying social contact is close or 
ordinary. Empirical evidence~\cite{Vollmer2001,Pierce2013,Skinner2015} 
suggests that the transmission associated with a close contact contributes
to reinforcement, but that with an ordinary contact does not. 
For a close contact, there is a memory effect in that the number $m$ of times
that a susceptible node receives the information from the adopted neighbors
is stored and used to determine the probability that the node actually 
adopts the behavior~\cite{Wang2015}, making the underlying dynamics 
non-Markovian. However, for an ordinary contact, a susceptible individual 
would accept the behavior from any adopted neighbor with probability $p$.  
We subsequently extend our model to empirical networks.

The dynamical process of social contagion can be described, as follows.
Initially, $E_{c}=E\mu$ edges are randomly selected as close contacts, associated
with which is reinforcement. $\mu$ denotes the probability 
that an edge is assigned as a closed contact. The remaining edges represent ordinary contacts. 
A contagion is initiated within a single adopted cluster of size $N\rho_{0}$, 
while the remaining nodes are in the susceptible state. The spreading process 
starts and proceeds according to the SAR model, and the nodal dynamical states 
evolve using the synchronous updating rule. Specifically, an individual ($i$) 
who has adopted the behavior attempts to transmit the behavioral information 
to its susceptible neighbors ($j$). For a close social contact, the 
transmission from the adopted end to the susceptible end occurs with 
probability $q$, and $j$ will successfully adopt the message or behavior 
only if it has received the information at least $T$ times, where $T$ 
is the adoption threshold. In this case, there is reinforcement. For an 
ordinary contact, $j$ becomes adopted with probability $p$, or the 
transmission rate, without contributing to the accumulated times $m$. The 
spreading process is repeated until all adopted individuals become extinct 
in the network and the dynamics have reached a stable steady state. 

For simplicity, we assume in our study that the recovery probability is 
$r=1$, i.e., an individual who adopts the behavior at time $t$ will 
transmit the behavior to all its susceptible neighbors at time $t+1$,
after which it will lose interest in the behavior and will not transmit 
the message or behavior again. In all cases, we set $q=1.0$.

\section{Theory} \label{sec:theory}

We present a theoretical analysis to elucidate the roles of close and 
ordinary contacts in social contagion dynamics based on the edge-based 
compartmental theory~\cite{Miller2012,Miller2013,Miller2014,Wang2017}. 
In our model, a node $u$ in a {\em cavity} state cannot transmit the 
behavioral information to its neighbors, but can receive it from 
others~\cite{Karrer2010}. The dynamical correlation among the states of the 
neighbors is characterized by the two types of the contacts. If node $u$ 
with degree $k_u$ is susceptible at time $t$, it does not belong to the 
cluster of initial seed of contagion and it receives the information less 
than $T$ times from close contacts. We write $k_u=k_u^C+k_u^O$, where $k_u^C$ 
and $k_u^O$ are the numbers of close and ordinary contacts of $u$, 
respectively. Let $\theta_X(t)$ ($X\in\{C,~O\}$) be the probability that, 
up to time $t$, the message or behavioral information has not been 
successfully transmitted from one end of one close or ordinary contact of 
$v$ to the other end $u$. Combining the two conditions, we obtain the 
probability that node $u$ is in the susceptible state as
\begin{widetext}
\begin{equation} \label{eq:S_K}
s(k,t)=(1-\rho_0)\sum_{k_C=0}^{k}{k \choose k_C}\mu^k(1.0-\mu)^{k-k_C} 
\sum_{m_C=0}^{T-1}\phi_{m_C}^C(k_C,t)\theta_O^{k-K_C},
\end{equation}
\end{widetext}
where $\phi_{m_C}^C$ is the probability that node $u$ has received $m_C$ 
pieces of information from close contacts by time $t$. 
The term $\sum_{k_C=0}^{k}{k \choose k_C}\mu^k(1.0-\mu)^{k-K_C}$ represents 
the probability that node $u$ has $k_C$ edges corresponding to close contacts. 
The detailed expression of $\phi_{m_X}^X$ is given by
\begin{equation}\label{eq:phi_A}
\phi_{m_X}^X(k_X,t)={k_X \choose m_X}\theta_X^{k_X-m_X}(1-\theta_X)^{m_X}.
\end{equation}
The fraction of susceptible nodes is 
\begin{equation} \label{eq:S}
S(t)=\sum_{k}p(k)s(k,t).
\end{equation}
We wish to obtain the expression of $\theta_X(t)$ according to its 
definition of $\theta_X(t)$. In our model, an edge of a susceptible 
individual $u$ can connect to a susceptible, an adopted or a recovered 
neighbor $v$. Accordingly, $\theta_X(t)$ consists of three parts: 
\begin{equation} \label{eq:theta_A}
\theta_X(t)=\xi_S^X(t)+\xi_A^X(t)+\xi_R^X(t),
\end{equation}
where $\xi_S^X(t)$, $\xi_A^X(t)$ and $\xi_R^X(t)$ are the probabilities that 
the neighbor is in the susceptible, adopted and recovered state, respectively, 
and the information or message has not been successfully transmitted to $u$ 
up to time $t$.

If $v$ is in the susceptible state, it cannot transmit the behavioral 
information or message to $u$. Moreover, because node $u$ is in the cavity 
state, node $v$ can only obtain the information or message from other 
neighbors excluding $u$. Up to time $t$, node $v$ with degree $k_v=k_C+k_O$ 
can obtain $m_x$ pieces of information from contacts of type $X$ with 
the probability
\begin{equation} \label{eq:phi_A_Nei}
\tau_{m_X}^X(k_X,t)={k_X-1 \choose m_X}\theta_X^{k_X-m_X-1}(1-\theta_X)^{m_X}.
\end{equation}
Node $v$ will remain in the susceptible state if it has not received 
more than $T$ pieces of message or information through close contacts, 
neither has it received any message from any ordinary contact. There are 
two cases because the contact between $v$ and $u$ is either close or 
ordinary. Combining Eqs.~(\ref{eq:phi_A}) and (\ref{eq:phi_A_Nei}), we
obtain the probability that node $v$ is in the susceptible state in 
each case as  
\begin{widetext}
\begin{eqnarray} \label{eq:S_k_A_Nei}
\Theta_C(k,t)& = & \sum_{k_C=0}^{k}{k \choose k_C}\mu^k(1.0-\mu)^{k-k_C}\sum_{m_C=0}^{T-1}\tau_{m_C}^C(k_C,t)\theta_O^{k-k_C},\\
\Theta_O(k,t)& = & \sum_{k_O=0}^{k}{k \choose k_O}\mu^{k-k_O}(1.0-\mu)^{k_O}\sum_{m_C=0}^{T-1}\theta_O^{k_O-1}\phi_{m_C}^C(k_C,t).
\end{eqnarray}
\end{widetext}
From the degree distribution $p(k)$, we obtain the probability that the edge 
of type $X$ is connected to a susceptible neighbor as
\begin{equation} \label{eq:xi_S_A}
\xi_S^X(t)=(1-\rho_0)\frac{\sum_{k}k p(k)\Theta_X(k,t)}{\langle k\rangle},
\end{equation}
where $kp(k)/\langle k\rangle$ is the probability that an edge is connected 
to a neighbor of degree $k$ in an uncorrelated network. 

If $v$ is an adopted node, it would transmit the information or message to 
a susceptible neighbor through an edge with probability $\lambda_X\in\{p, q\}$,
leading to a decrease in $\theta_X(t)$. We thus have
\begin{equation} \label{eq:D_theta_A}
\frac{d\theta_X}{dt}=-\lambda_X\xi_A^X.
\end{equation}
Otherwise, $v$ fails to transmit the information or message to neighbor and
becomes recovered, $\xi_R^X$ will increase. This means that the increment 
$d\xi_R^X$ must satisfy two conditions: (a) the adopted neighbor has not 
transmitted the behavioral information or message to neighbors, which occurs
with the probability $1-\lambda_X$, and (b) the adopted neighbor recovers with 
probability $\gamma$. We thus have
\begin{equation}\label{eq:xi_R_A}
\frac{d\xi_R^X}{dt}=\gamma(1-\lambda_X)\xi_A^X.
\end{equation}
From Eqs.~(\ref{eq:D_theta_A}) and (\ref{eq:xi_R_A}) as well as the
initial conditions $\theta_X(0)=1$ and $\xi_R^X(0)=0$, we obtain
\begin{equation} \label{eq:xi_R}
\xi_R^X(t)=\gamma[1-\theta_X(t)](1-\lambda_X)/\lambda_X.
\end{equation}
Combining Eqs.~(\ref{eq:theta_A}), (\ref{eq:xi_S_A}), (\ref{eq:D_theta_A}), 
and (\ref{eq:xi_R}), we obtain the following equation that governs the 
evolution of $\theta_X(t)$: 
\begin{eqnarray} \label{eq:thetatime}
\frac{d\theta_X(t)}{dt}& = & (1-\rho_0)\lambda_X\frac{\sum_{k}kp(k)\Theta_X(k,t)}{\langle k\rangle} \\ \nonumber
& + & \gamma[1-\theta_X(t)](1-\lambda_X)-\lambda_X\theta_X(t).
\end{eqnarray}
Using the identity $S(t)+A(t)+R(t)=1$ and the fact that the rate $dA(t)/dt$ 
is equal to the rate at which $S(t)$ decreases, we have
\begin{eqnarray}
\frac{dA(t)}{dt} & = & -\frac{dS(t)}{dt}-\gamma A(t), \\ \label{eq:evolutionar1}
\frac{dR(t)}{dt} & = & \gamma A(t), \label{eq:evolutionar2}
\end{eqnarray}  
where $A(t)$ and $R(t)$ are the fractions of the adopted and recovered 
population at time $t$, respectively. Accordingly, $R(\infty)$ represents the
final fraction of the recovered population after the system has reached a 
steady state that no longer has any adopted node.
Equations~(\ref{eq:S_K}-\ref{eq:S}) and 
(\ref{eq:thetatime}-\ref{eq:evolutionar2}) describe a full and general 
picture of the contagion dynamics.

The steady state fraction of nodes that have adopted the behavior can be 
obtained from Eq.~(\ref{eq:thetatime}) by taking the limit 
$t\rightarrow\infty$ as
\begin{eqnarray} \label{eq:thetax}
\theta_X(\infty) & = & (1-\rho_0)\frac{\sum_{k}kp(k)\Theta_X(k,\infty)}
{\langle k\rangle}\\ \nonumber
& + & \gamma[1-\theta_X(\infty)]\frac{1-\lambda_X}{\lambda_X}.
\end{eqnarray}
It is now feasible to analyze the critical information transmission 
probability. In the presence of social reinforcement ($T>1$), a vanishingly 
small number of seeds cannot trigger a global adoption. It is useful to 
consider a finite fraction of seeds: $\rho_0>0$. In this case, 
$\theta_X(\infty)=1$ is not a solution of Eq.~(\ref{eq:thetax}). At the 
critical point of first-order phase transition, the condition
\begin{equation} \label{eq:firstorder}
\frac{\partial \theta_C(\theta_C(\infty),\theta_O(\infty))}{\partial
\theta_O(\infty))} \frac{\partial \theta_O(\theta_C(\infty),\theta_O(\infty))}
{\partial \theta_C(\infty))}=1
\end{equation}
is fulfilled~\cite{Parshani2010}, where $\theta_X$ 
$[\theta_C(\infty),\theta_O(\infty)]$ is the right-hand side of 
Eq.~(\ref{eq:thetax}).

\section{Results} \label{sec:results}

\subsection{Erd\"{o}s-R\'{e}nyi (ER) networks} \label{subsec:ernetwork}

We simulate the spreading dynamics on ER networks with $N_r=500$ independent 
realizations. A new random network with the same rewiring probability $p_e$ 
is built after every 25 independent realizations of the spreading dynamics. 
Unless otherwise specified, the simulation parameter values in most cases 
are $N=10^{4}$, $\rho_0=0.003$, $q=1.0$ and $\langle k\rangle=10.0$ 
(corresponding to $p_{e}=0.001$). The number of edges is thus 
$E=5\times 10^{4}$. For small values of $\rho_{0}$, analytic prediction 
can be obtained to uncover the roles of the two distinct types of social 
contacts in contagion dynamics.

\begin{figure*}[!ht]
\centering
\includegraphics[width=0.9\linewidth]{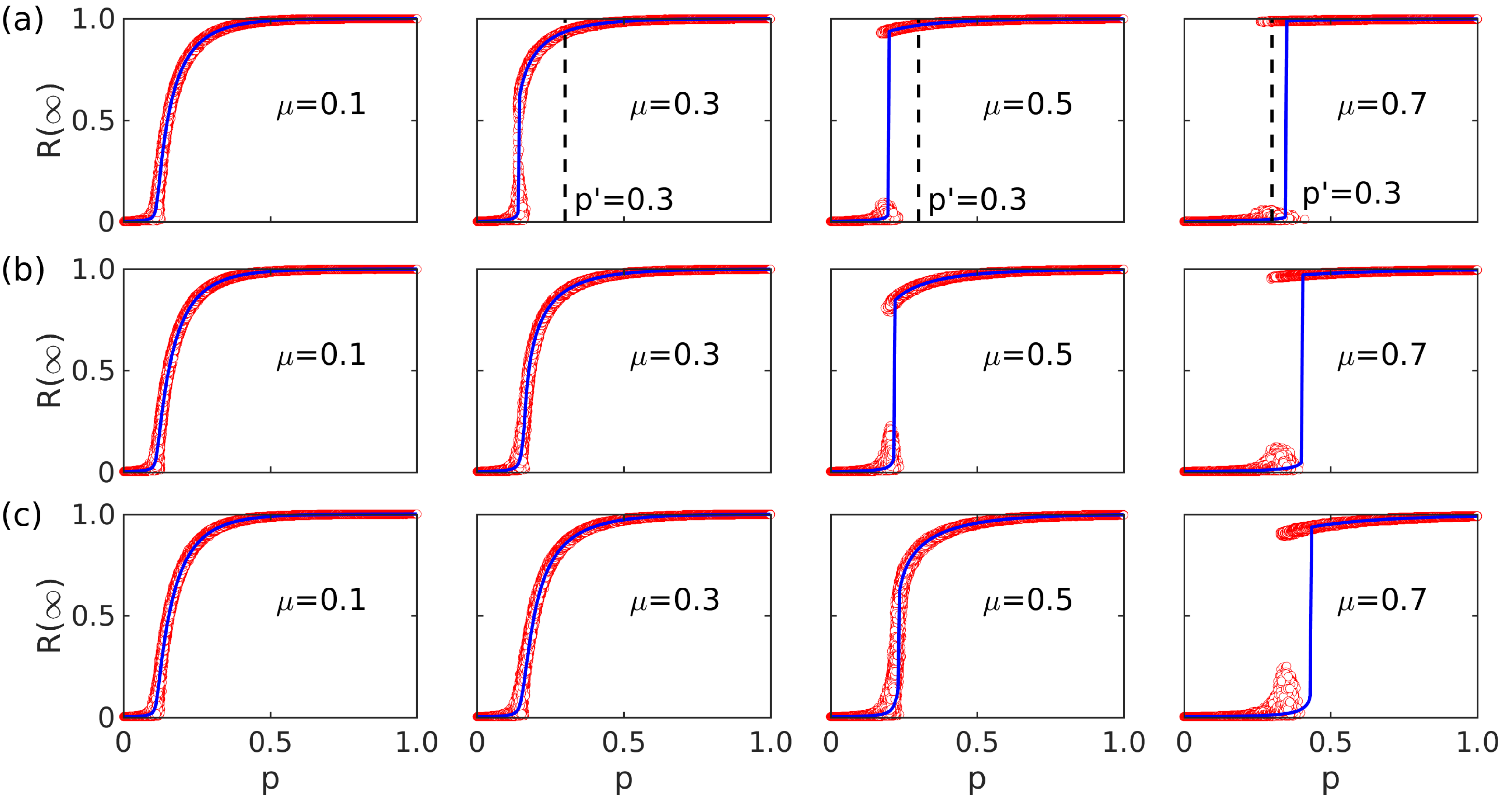}
\caption{ (Color online) Numerically obtained and theoretically predicted final 
fraction of the recovered population for different parameter values.
Simulation results (red circles) and theoretical prediction of
Eq.~(\ref{eq:thetatime}) (blue solid lines) for the final
recovered population versus transmission rate $p$ for three different values of $T$: (a) $T=3$, (b) $T=4$, and 
(c) $T=5$, where for each value of $T$, the results for four different 
values of $\mu$ are presented. Each data point is obtained from a single 
realization. The black vertical dashed lines indicating the position of 
$p^{\prime}=0.3$ are included in (a) for corresponding the results to the 
dynamical behaviors of the system in Fig.~\ref{fig:transmission1}.}
\label{fig:atfmu}
\end{figure*}

A fundamental feature of the spreading dynamics of social contagion is
the emergence of phase transitions: as a system parameter varies through
a critical point, an outbreak occurs. The transition can be continuous
(second order) or discontinuous (first order). To gain insights into the 
role of close versus ordinary contacts in the characteristics of phase 
transition, we calculate the final fraction of the recovered population
for different values of $T$ and $\mu$ numerically and theoretically, as 
shown in Fig.~\ref{fig:atfmu}.  
In all cases, there is a phase transition. As the fraction of close 
contacts in the network is increased, there is a change in the nature of 
the phase transition from being continuous to discontinuous, where the 
abrupt change in the final recovered population at the transition point
can be quite large for relatively large values of the fraction. The dynamical
origin of the discontinuity can be attributed to the relative abundance of 
the close contacts that lay the ground for the occurrence of a drastic 
avalanche type of process, as suggested by the previous result that 
a sufficient number of close contacts with a relatively large transmission 
rate can trigger a massive outbreak of contagion~\cite{Wang2015,Cui2018}. 
For relatively higher values of the adoption threshold $T$, second order 
transitions are more likely, as illustrated in Fig.~\ref{fig:atfmu}(c). 
The reason lies in the susceptible individuals in the critical state, i.e., 
the individuals who have been informed by their adopted neighbors through 
close contacts but have not adopted the message or information yet. For 
a large value of $T$, it is more difficult to make these
individuals adopt the behavior, thereby requiring more transmission to 
trigger a massive adoption process. The contribution of reinforcement 
in this case is insignificant, leading to a smaller avalanche 
size~\cite{Ausloos2014,Cui2018}. Note the good agreement between numerical 
and theoretical results, which validates the edge-based compartmental 
approach. 

Figure~\ref{fig:atfmu} also shows that close contacts of sufficient 
number tend to delay the outbreak of massive contagion, as
the corresponding threshold values are larger than 
$T_{C}=\langle k\rangle/(\langle k^2\rangle-\langle k\rangle)
=\langle k\rangle^{-1}$ as predicted by the bond-percolation theory for 
the conventional susceptible-infected-refractory (SIR) 
dynamics~\cite{Callaway2000,Pastor2015}. More close contacts are more
effective at preventing a large scale outbreak, regardless of the values of
the transmission rate and adoption threshold $T$. This is quite surprising
as previous work had concluded that dense contacts with social reinforcement 
would in general facilitate contagion~\cite{Centola2010}.

\begin{figure*}[!ht]
\centering
\includegraphics[width=0.9\linewidth]{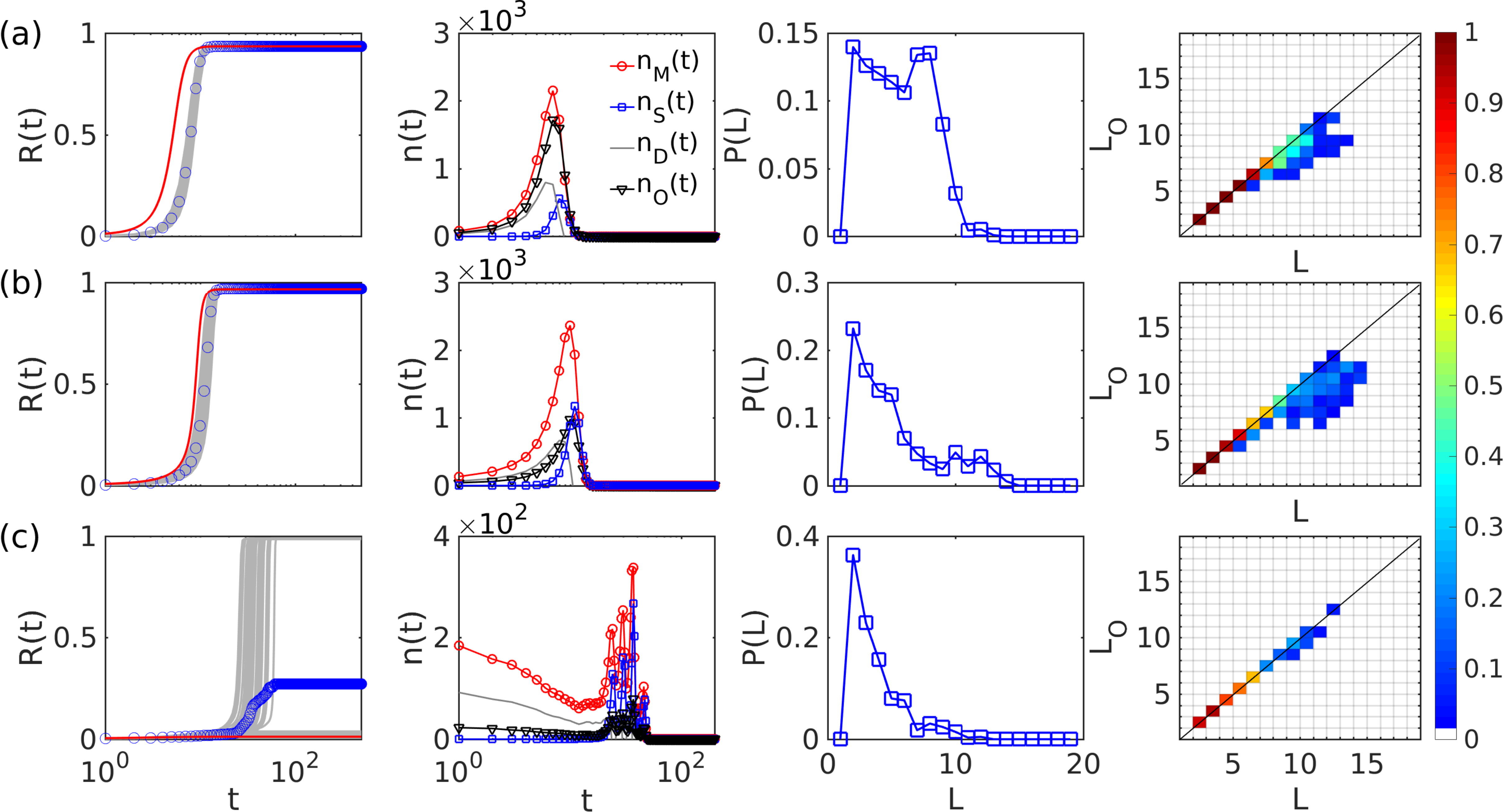}
\caption{ (Color online) Understanding the effects of ordinary contacts on 
spreading dynamics. Four key statistical characterizing quantities are
shown for $p^\prime=0.3$, $T = 3$ and $q = 1.0$: the fraction of recovered
population (the first column), the numbers of three types of transmission
events (the second column), the distribution $P(L)$ of diffusion path
lengths $L$ (the third column), and the frequency spectrum of the
occurrence of the ordinary transmission events in the various diffusion paths 
(the fourth column). Three different values of the parameter $\mu$ are used:
(a) $\mu=0.3$, (b) $\mu=0.5$ and (c) $\mu=0.7$. 
The red solid lines in the first column are the analytical predictions, the 
blue circles represent the fractions of the recovered populations obtained by 
averaging $N_r$ independent realizations, and each light gray solid curve 
in each panel in the left column corresponds to one independent realization.
The position of $p^{\prime}=0.3$ has also been specified in 
Fig.~\ref{fig:atfmu}.}
\label{fig:transmission1}
\end{figure*}

The counterintuitive phenomenon can be understood by estimating the 
contributions of the two types of distinct contacts to the spreading 
dynamics. Figure~\ref{fig:transmission1} shows the key statistical 
characterizing quantities for $p^\prime=0.3$: the fraction of recovered 
population (the first column), the numbers of three types of transmission 
events (the second column), the distribution $P(L)$ of diffusion path lengths 
$L$ (the third column), and the frequency spectrum of the occurrence of  
ordinary transmission events in the various diffusion paths (the 4th column). 
More specifically, in our model, transmission events can be classified into 
three major categories: (a) transmission associated with ordinary contacts 
which occurs with probability $p$ - ordinary transmission, (b) transmission 
along close contacts which is able to drive the individuals to successfully 
receive the information about the behavior for $m_{i}(t)<T$ - intermediate 
reinforced transmission, and (c) transmission associated close contacts
which results in acceptance of the behavior when the condition $m_{i}(t)=T$ 
is met - successful reinforced transmission, which occurs with probability 
$q$. The numbers of the three types of transmission events are denoted as
$n_O(t)$, $n_M(t)$, $n_S(t)$, respectively. The quantity
$n_{D}(t)=n_M(t)/(T-1)-n_S(t)$ represents the required additional 
minimum number of transmission events to stimulate all the 
remaining susceptible individuals in the critical state, after stimulation 
from successful reinforced transmission. About the statistical distribution
$P(L)$, we note that a diffusion path of length $L$ is a combination of 
$L_{S}$ successfully reinforced transmission events and $L_{O}$ ordinary 
transmission events: $L=L_{S}+L_{O}$. With respect to the frequencies of 
occurrence of the ordinary transmission events along various diffusion paths,
each color square at the position $(L^{\prime},L^{\prime}_{O})$ of the 
spectrum represents the probability that there are $L^{\prime}_{O}$ ordinary 
transmission events along the diffusion path of length $L^{\prime}$, where the
non-empty squares below the diagonal lines indicate that ordinary 
transmission events are engaged in propagating the contagion forward 
together with the successful reinforced transmission events. 

Figure~\ref{fig:transmission1} illustrates a key feature of the system:
when the number of ordinary transmission events approximately matches 
$n_{D}(t)$ during the spreading process, i.e., $n_{O}(t)\approx n_{D}(t)$,
the system evolves into the state of maximum contagion. For a relatively
small value of $\mu$ [Fig.~\ref{fig:transmission1}(a)], close contacts are insufficient 
so that the amount of susceptible individuals in critical state are less than 
ordinary transmissions ought to stimulate. 
For a larger value of $\mu$, ordinary transmission 
events are rare and the two types of reinforced transmission events 
decrease in number even for a fixed value of the transmission rate $p$.
The third column of Fig.~\ref{fig:transmission1} indicates the existence
of an appreciable number of diffusion paths of intermediate and long length
(e.g., $L\geq 7$) for the case where contagion is prevalent 
[Figs.~\ref{fig:transmission1}(a,b)], and that short paths dominate when contagion is prohibited
[Fig.~\ref{fig:transmission1}(c)]. The behavior of the recovered population 
size suggests that long diffusion paths favor contagion when 
the value of the transmission rate and the fraction of ordinary contacts
are proper to enable the dynamics. 

Which elements contribute to the long paths and how do they facilitate 
contagion? The fourth columns of Fig.~\ref{fig:transmission1} 
provide a partial answer. For the cases shown, short paths ($L\leq 5$) are 
mostly due to ordinary transmission events, while ordinary and successful 
reinforced transmission events have different contributions to longer 
diffusion paths for different values of $\mu$. In the fourth column of 
Figs.~\ref{fig:transmission1}(a,b), the non-empty squares below the  
diagonal lines have a relatively wide distribution, indicating that 
ordinary transmission tends to ``cooperate'' with successful reinforced 
transmission to propagate the contagion forward together, enabling 
persistent transmission and generating long diffusion paths.
Specifically, the more frequently ordinary transmission events are engaged 
in contagion propagation along longer paths (more non-empty squares below 
the diagonal lines in regions of large values of $L$), the population
has more adopted individuals. That is, the existence of the long diffusion 
paths containing a substantial number of ordinary transmission events imply
a long-time cooperative relationship between the two different types of
transmission events. The extent of contagion is maximized when
the number of susceptible individuals in the critical state induced by close 
contacts matches the number of ordinary transmission events as induced by
ordinary contacts. There then exists an optimal fraction of close contacts 
to facilitate contagion to the maximum extent.  

Note that, in the first column of Fig.~\ref{fig:transmission1}(c), there
is a discrepancy between the analytical and numerical results, especially 
near the critical point. This is due to the fluctuations as the system evolves 
from a local contagion state to a global prevalent state, as indicated by
the spread in the gray curves from individual realizations.

\begin{figure*}[!ht]
\centering
\includegraphics[width=\linewidth]{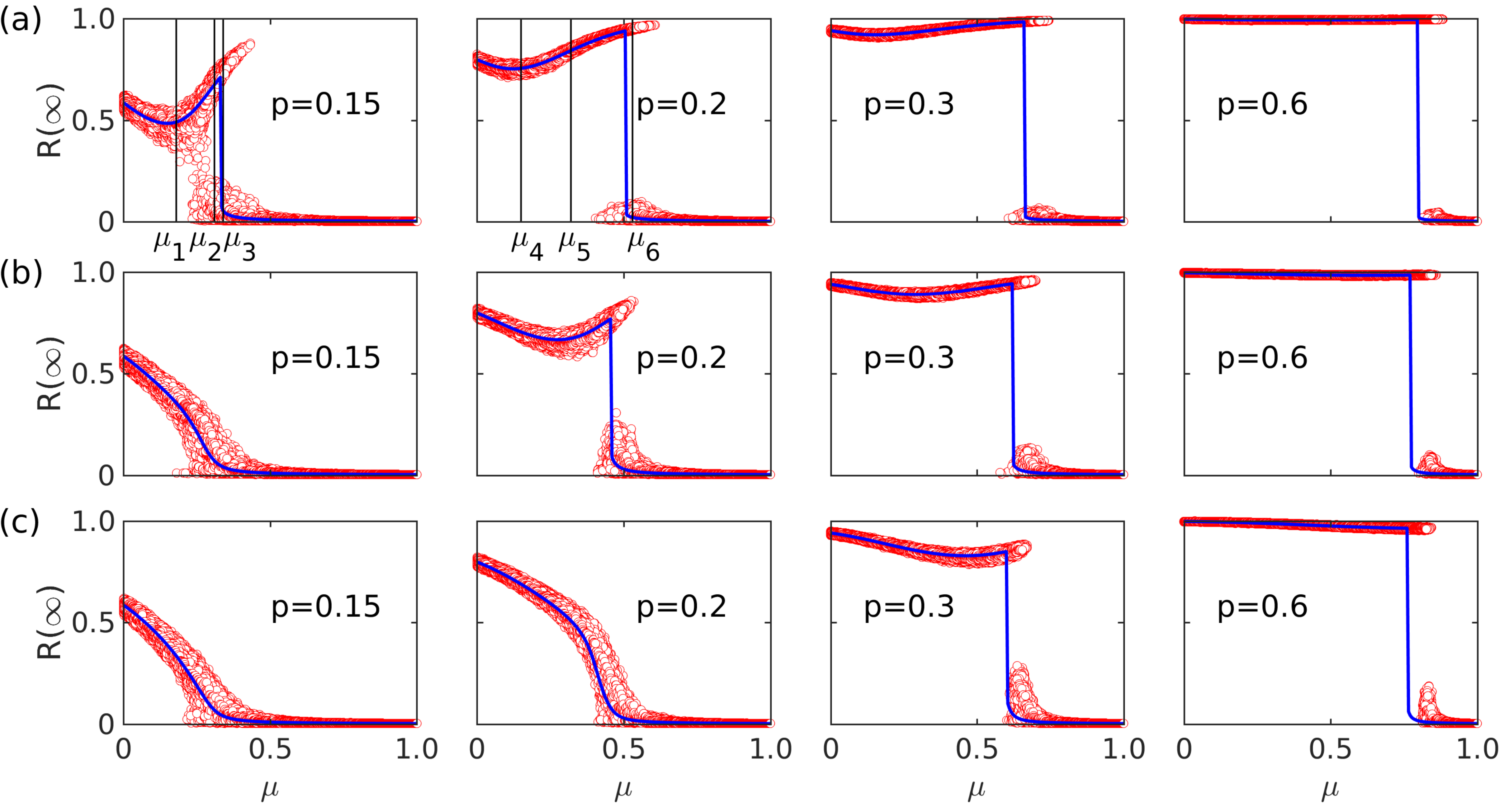}
\caption{ (Color online) Existence of an optimal fraction of close transmission
events for global spreading. Shown is the final recovered population 
as a function of $\mu$ for three different values of $T$ for $q=1.0$: 
(a) $T=3$, (b) $T=4$ and (c) $T=5$. For each value of $T$, simulation 
results (red circles) and analytical predictions (blue solid lines) are 
shown for four different values of $p$. Two groups of vertical solid 
lines ($\mu_1$, $\mu_2$, $\mu_3$) and ($\mu_4$, $\mu_5$, $\mu_6$) are 
indicated in (a), corresponding to Figs.~\ref{fig:optimalmu} and 
\ref{fig:optimalmu2}, respectively.}
\label{fig:atfp}
\end{figure*}

The existence of an optimal fraction of close transmission events can
be further illustrated numerically and analytically, as shown in 
Fig.~\ref{fig:atfp}, where the existence of the optimal fraction is 
more apparent for small values of $p$ and $T$. This means that the system 
can be harnessed to reach a maximum level of contagion without the need to 
change the transmission rate $p$. For a relatively large value of $p$ and
a not too large value of $\mu$, ordinary contacts can be exploited to 
maximize the spreading in that it is more likely that the interplay between
ordinary and successful reinforced transmission events will generate 
long diffusion paths. Note that the transitions with respect to varying
$\mu$ are mostly discontinuous, especially in cases where there exists 
an interval of $\mu$ values with maximum spreading.

\begin{figure*}[!ht]
\centering
\includegraphics[width=\linewidth]{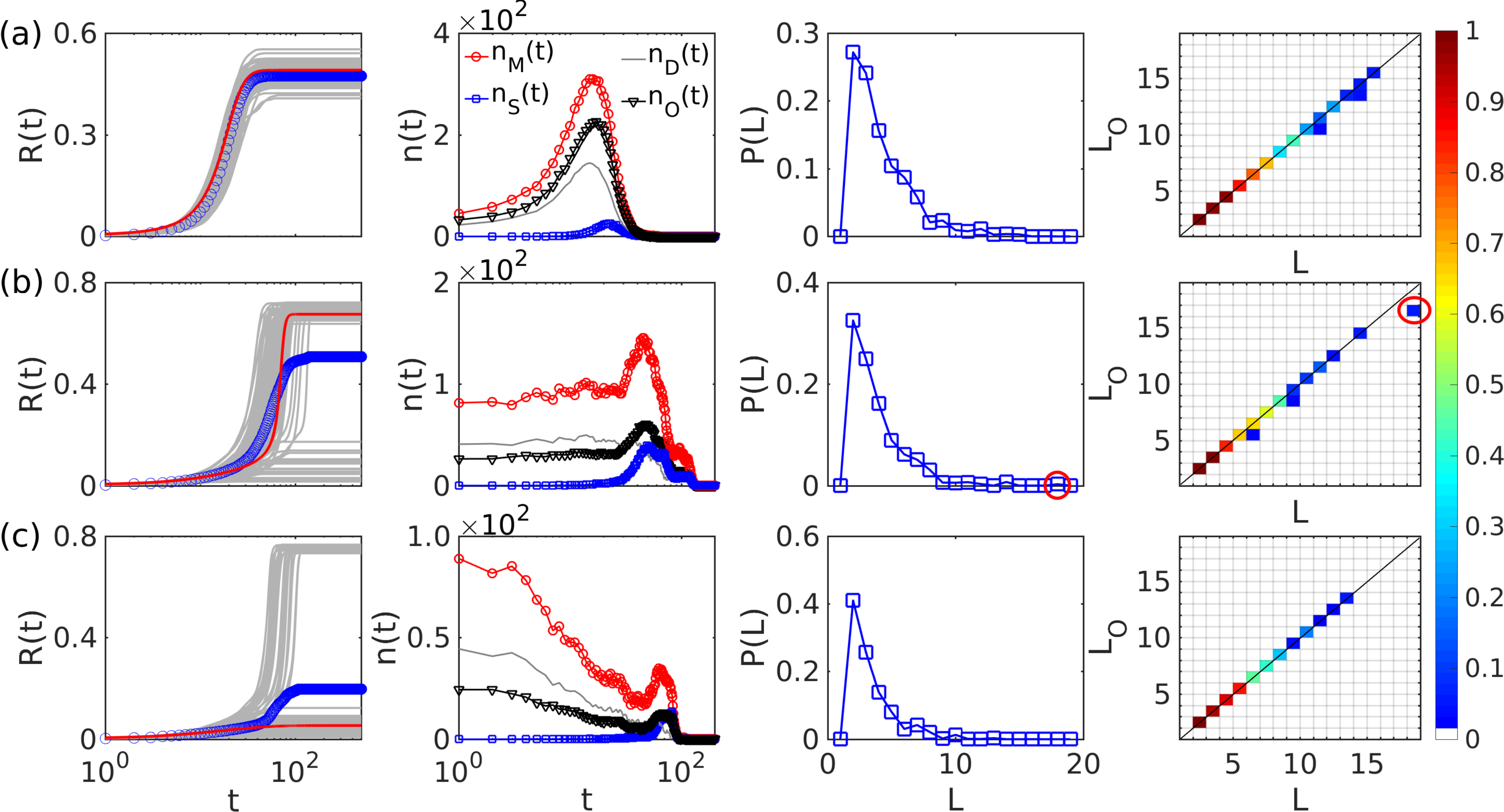}
\caption{ (Color online) Illustrations of the role of ordinary contacts 
in promoting spreading. Shown are the behaviors of four
statistical characterizing quantities for different values of $\mu_o$:   
(a) $\mu_{o}>\mu_1=0.18$, (b) $\mu_{o}\approx\mu_2=0.31$ and 
(c) $\mu_{o}<\mu_3=0.34$, the positions of which are also labeled in 
Fig.~\ref{fig:atfp}. Other parameter values are $p=0.15$ and $q=1.0$.}
\label{fig:optimalmu}
\end{figure*}

Figure~\ref{fig:optimalmu} provides a visual picture of the results in 
Fig.~\ref{fig:atfp}, which further validates the physical picture for 
achieving maximum contagion in Fig.~\ref{fig:transmission1} through 
a match between the numbers of close and ordinary transmission events: 
$n_{m}(t)\gg n_{e}(t)$ for $\mu_{2}=\mu_{o}=0.31$. Prevalence of 
the recovered population is not possible when close contacts dominate 
so there is a lack of cooperation with ordinary contacts, regardless of 
the occurrence of intermediate reinforced transmission [e.g., the second
plot in Fig.~\ref{fig:transmission1}(c)]. Overall, frequent cooperation 
between ordinary and successful reinforced transmission events along long 
diffusion paths is key to and plays a more important role in contagion 
prevalence rather than the mere existence of the long diffusion paths. 
Ordinary contacts are thus indispensable for successful social contagion.  

\begin{figure*}[!ht]
\centering
\includegraphics[width=\linewidth]{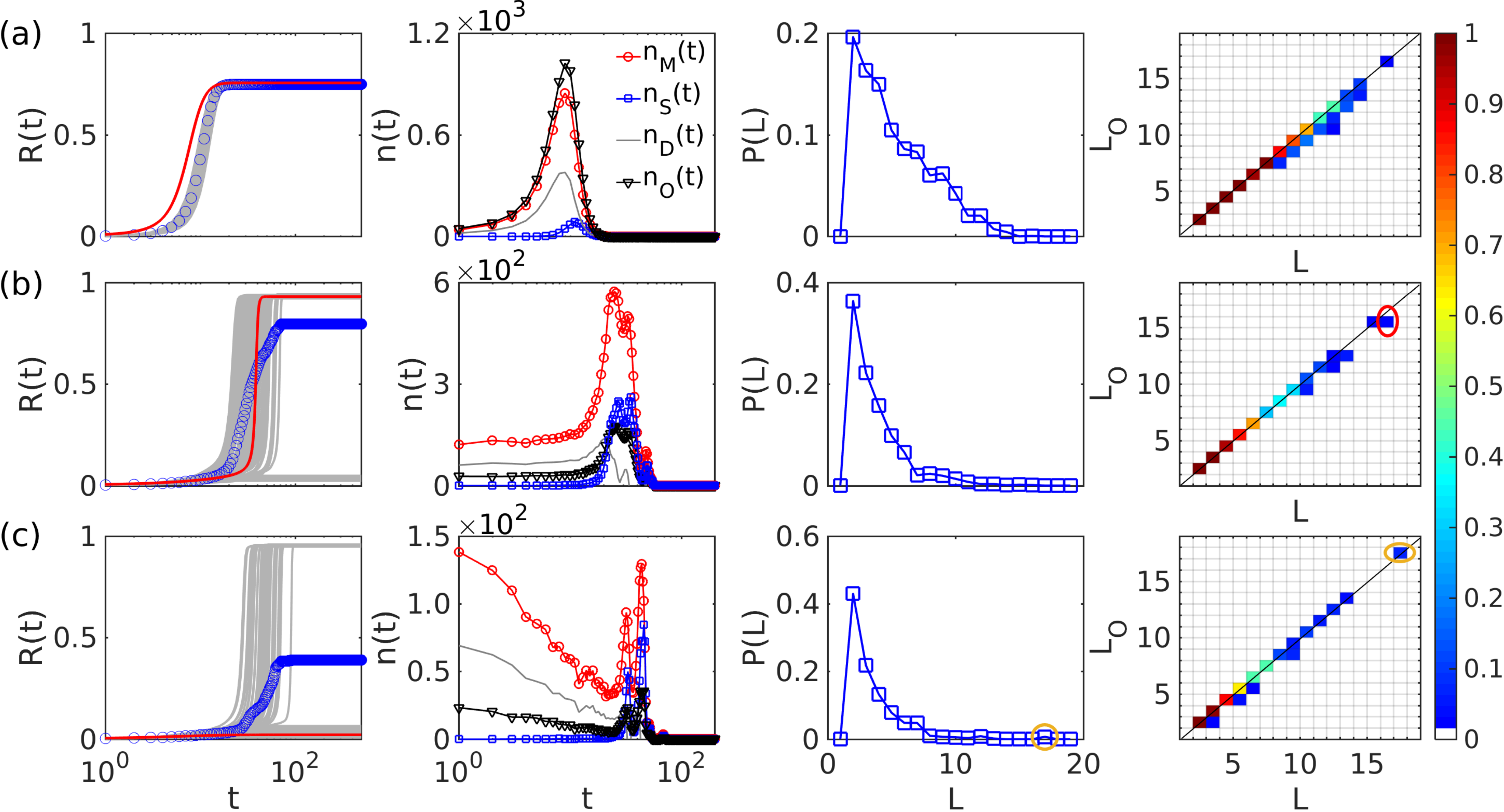}
\caption{ (Color online) Further illustrations of the role of ordinary 
contacts in promoting spreading. Shown are the behaviors of the four 
statistical characterizing quantities for (a) $\mu_4=0.15$, (b) $\mu_5=0.48$, 
and (c) $\mu_6=0.53$, corresponding to the three points of $\mu$ labeled in 
the 2nd panel of Fig.~\ref{fig:atfp}(a). Other parameters are $p=0.2$ 
and $q=1.0$.}
\label{fig:optimalmu2}
\end{figure*}

Further support for requiring a non-trivial interplay between close and 
ordinary contacts in promoting large scale spreading is provided in 
Fig.~\ref{fig:optimalmu2}, where the colored square in the red circle reveals 
the cooperation between close and ordinary contacts along a long diffusion 
path of length $L=16$ that consists of $15$ ordinary and one successfully 
reinforced transmission events. Note that, in this case, longer diffusion 
paths exist, e.g., a path of length $L=17$, which consist of only one 
type of transmission events for small contagion size (the third and 
fourth plots in Fig.~\ref{fig:optimalmu2}(c) - the symbols in the brown 
circles). However, these paths do not contribute to global spreading. 
The main reason for the discrepancies between analytical predictions 
and simulated results illustrated in the first columns of (b,c) in both 
Figs.~\ref{fig:optimalmu} and \ref{fig:optimalmu2} is again the increasing 
fluctuations of the system near the outbreak threshold. 

\begin{figure*}[!ht]
\centering
\includegraphics[width=\linewidth]{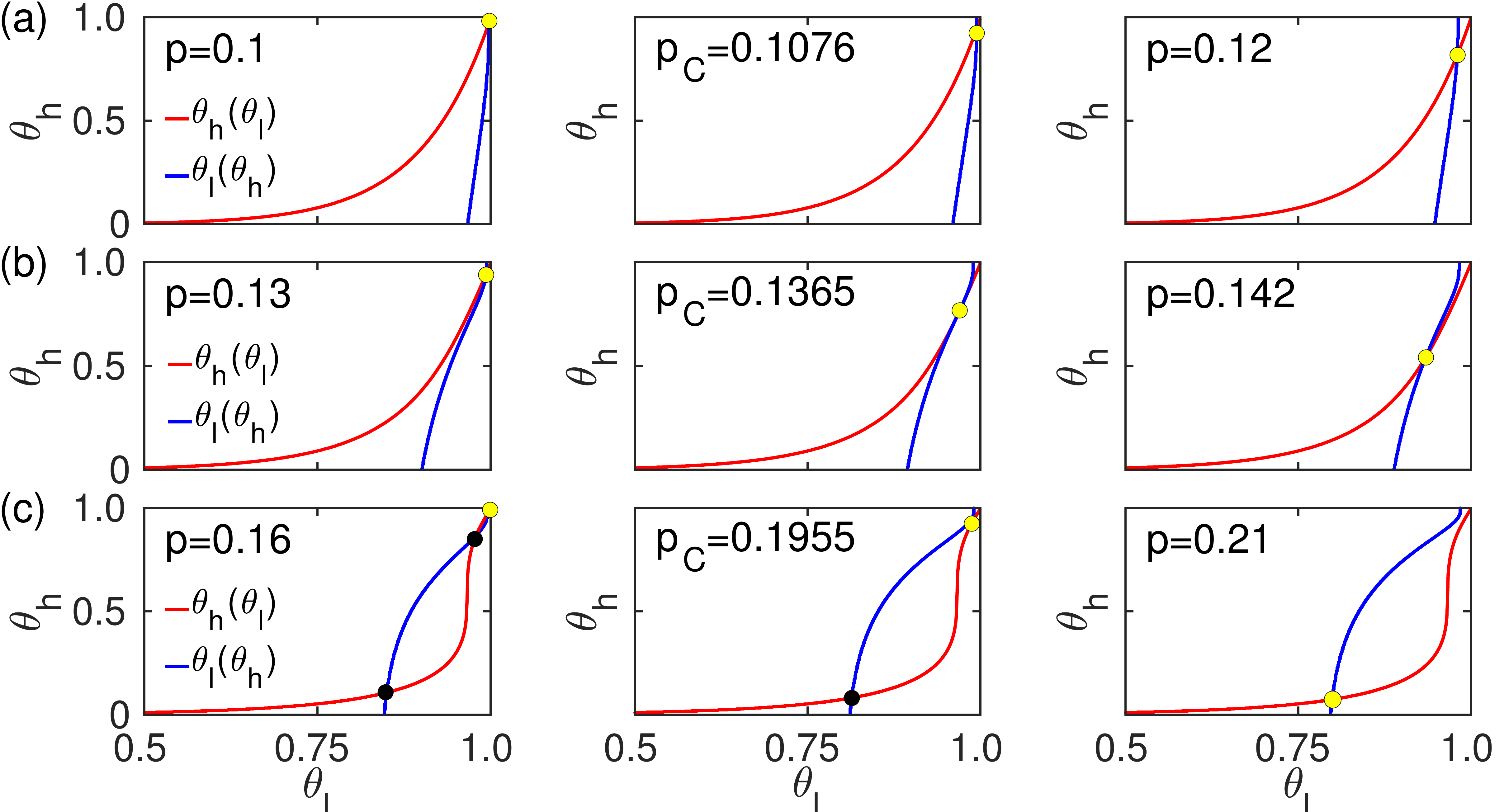}
\caption{ (Color online) Illustrations of graphical solutions of the 
bifurcation analysis for three cases: (a) continuous (second order) transition 
for $\mu=0.1$; (b) transition crossing the triple point for $\mu_C=0.245$; 
(c) discontinuous (first order) transition for $\mu=0.5$. Other parameters 
are $T=3$ and $q=1.0$.}
\label{fig:phasetransition}
\end{figure*}

The phase transition scenarios of the system can be graphically understood
through a bifurcation analysis of Eq.~(\ref{eq:thetax}), as shown
in Fig.~\ref{fig:phasetransition}. There are three distinct scenarios: 
continuous transition [Fig.~\ref{fig:phasetransition}(a)], transition at 
the triple point [Fig.~\ref{fig:phasetransition}(b)], and discontinuous 
transitions [Fig.~\ref{fig:phasetransition}(c)]. For the continuous 
transition, the graphical solution (the crossing point between the two 
relevant curves) moves gradually away from the position of $\theta_s=1$ 
and $\theta_n=1$ to that of $\theta_s<1$ and $\theta_n<1$. The crossing point 
begins to leave the top right corner of the plot at the critical point 
$p=p_C$. Associated with the triple point transition, although there 
exists one crossing point, two curves are close to each other near the 
crossing point that tends to move abruptly, which is indicative of a 
discontinuous transition. Note that the triple point can be estimated with 
the aid of pattern features of the solution lines, as illustrated in 
Fig.~\ref{fig:phasetransition}(b). For the discontinuous transitions, a 
solution for which the crossing points decrease from three (corresponding to 
contagion decay) to two (corresponding to contagion outbreak) until only 
one (contagion prevailing). Furthermore, Eq.~(\ref{eq:firstorder}) gives 
the tangent point of the two curves presented in the second subplot of 
Fig.~\ref{fig:phasetransition}(c). The crossing point (yellow solid circles) 
closest to the top right corner of the plot (i.e., $\theta_s=1$ and 
$\theta_n=1$) is physically meaningful, because the dynamics start from
the initial conditions $\theta_s\approx 1$ and $\theta_n\approx 1$. The 
movement of the yellow points is caused by the shift of blue lines, 
indicating again the role played by the transmission events due to 
ordinary contacts in the phase transition. 

\begin{figure*}[!ht]
\centering
\includegraphics[width=\linewidth]{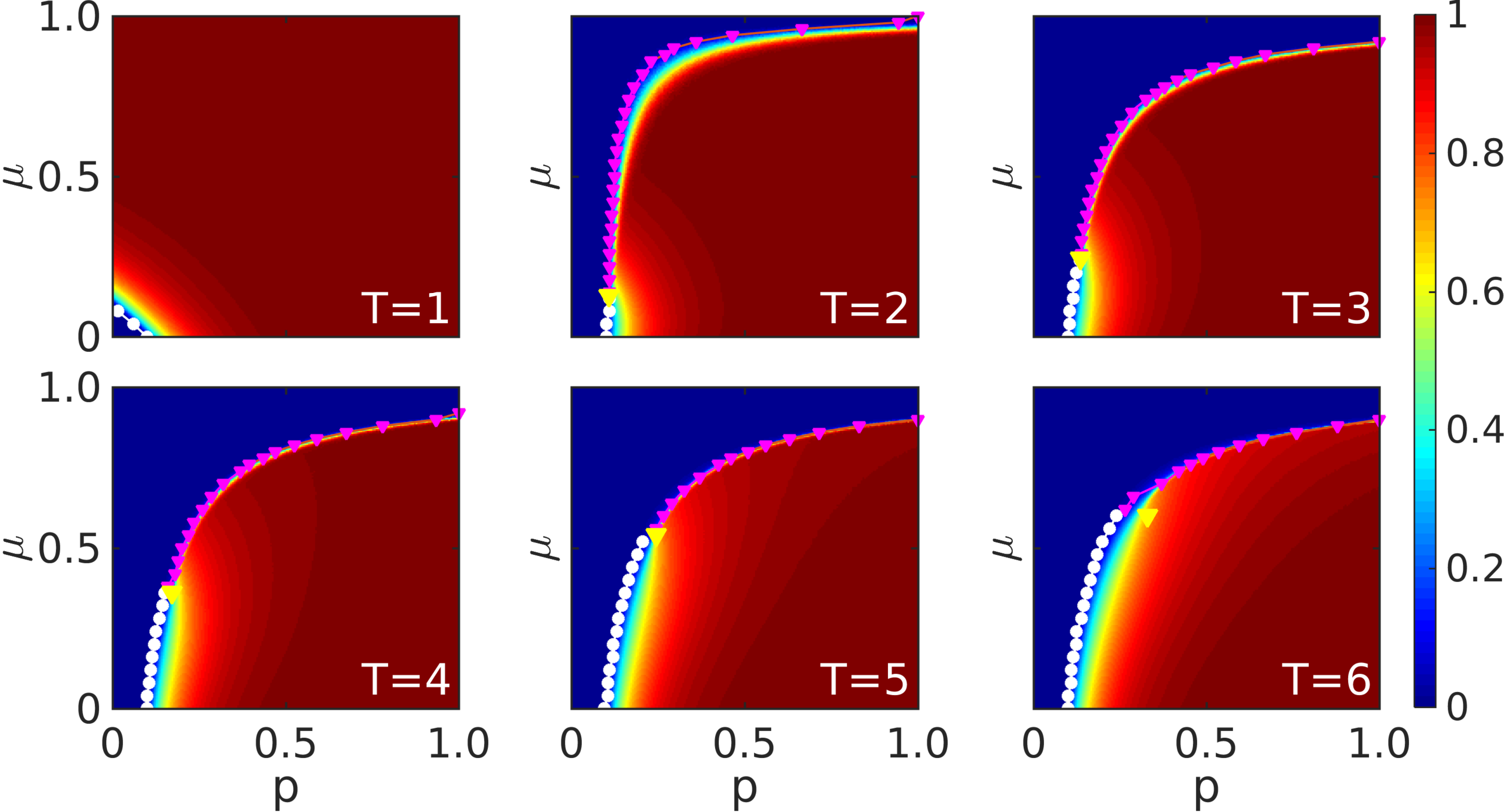}
\caption{ (Color online) Numerically obtained final stationary distributions 
of the recovered population. Shown are the distributions versus $p$ and $\mu$ 
for different values of $T$ and $q = 1.0$. The large yellow inverted 
triangle represents the theoretically estimated triple point, while other 
markers denote the numerically estimated phase boundaries in terms of 
the relative variance $\nu(\infty)$. The white circles (pink inverted 
triangles) represent the boundaries across which continuous (discontinuous) 
transitions occur.} 
\label{fig:simulation_all}
\end{figure*}

\begin{figure*}[!ht]
\centering
\includegraphics[width=\linewidth]{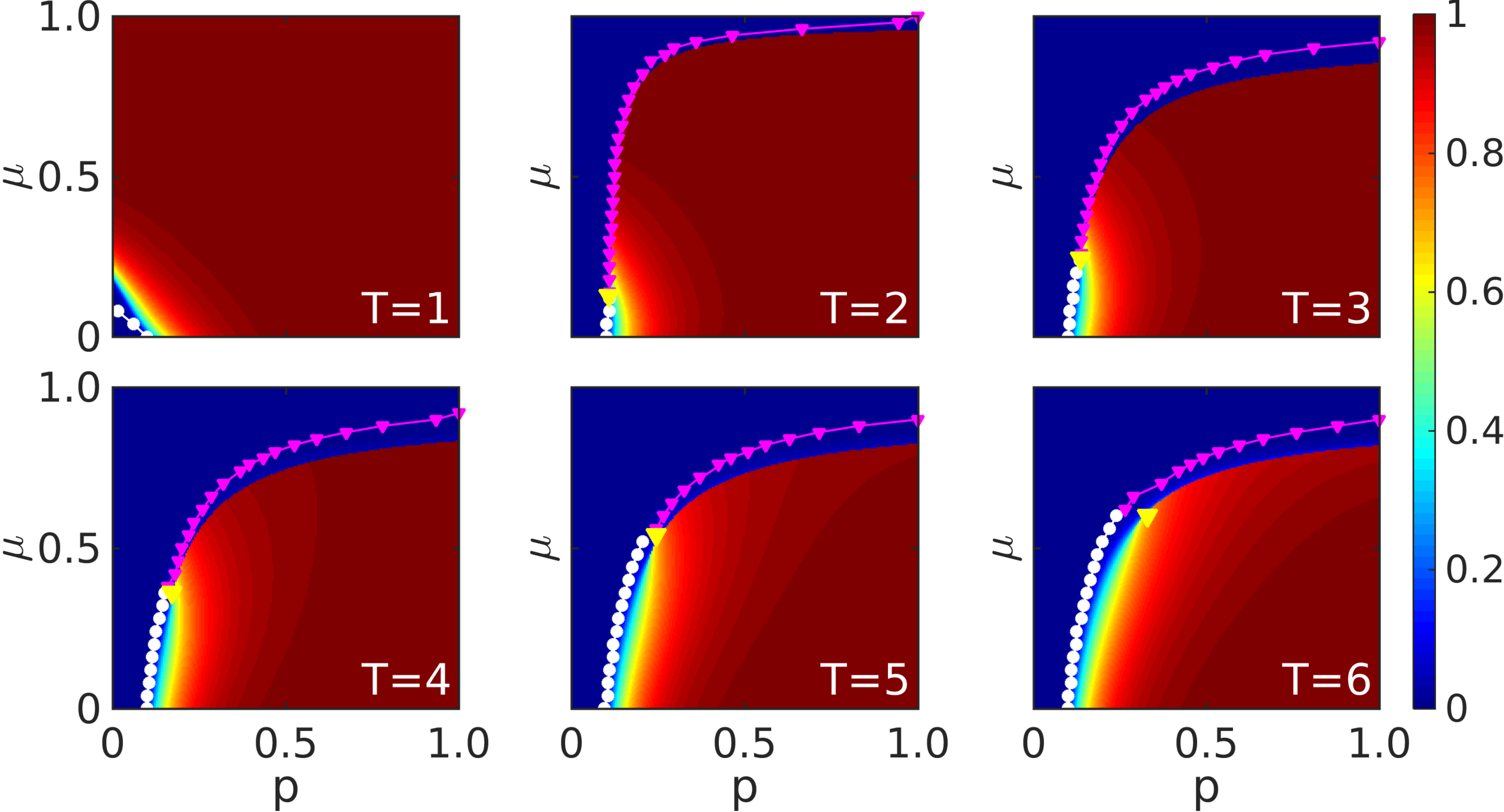}
\caption{ (Color online) Theoretically predicted final stationary 
distributions of recovered population. Legends are the same as those in 
Fig.~\ref{fig:simulation_all}. There is a good agreement between the 
numerical results in Fig.~\ref{fig:simulation_all} and the theoretical
predictions here.}
\label{fig:analysis_all}
\end{figure*}

Figures~\ref{fig:simulation_all} and \ref{fig:analysis_all} provide a 
comprehensive picture in the parameter plane of ($\mu$, $p$) of the 
spreading dynamics, where the former is numerically obtained while the 
latter is theoretical predictions. Note that the numerically identified 
critical boundaries and the analytically estimated triple point are 
included. Both figures represent evidence for the coexistence of continuous 
and discontinuous phase transitions separated by the triple point. More 
specifically, smooth color variations and dramatic differences in the 
color indicate continuous and discontinuous types of transition, 
respectively. There is an overall good agreement between the simulation 
results and the analytic predictions, although there is deviation between
the two types of results in the region of large values of $\mu$. The 
discrepancies are somewhat expected, due to the numerical uncertainties
in identifying the outbreak phase of the contagion through a finite number
of statistically independent realizations. In general, the analytical 
predictions are more reliable, as evidenced by the better agreement with the
numerical results for the limited parameter setting in which more extensive 
stochastic simulations are computationally feasible (e.g., the results in 
Figs.~\ref{fig:atfmu} and \ref{fig:atfp}). The results in 
Figs.~\ref{fig:simulation_all} and \ref{fig:analysis_all} indicate that 
the threshold and transition classes are predominately determined by 
the relative abundance of close contacts with respect to ordinary 
contacts in the underlying social network. Another feature is that the 
spreading dynamics in the presence of reinforcement ($T>1$) is completely 
suppressed when the value of $\mu$ exceeds certain threshold whose value
decreases with $T$, due to the lack of match between the two types of social
contacts. Note that, in Figs.~\ref{fig:simulation_all} and 
\ref{fig:analysis_all}, the boundaries of continuous transitions are longer 
for higher values of $T$, a result that is consistent with those in 
Fig.~\ref{fig:atfmu}.

\begin{figure*}[!ht]
\centering
\includegraphics[width=\linewidth]{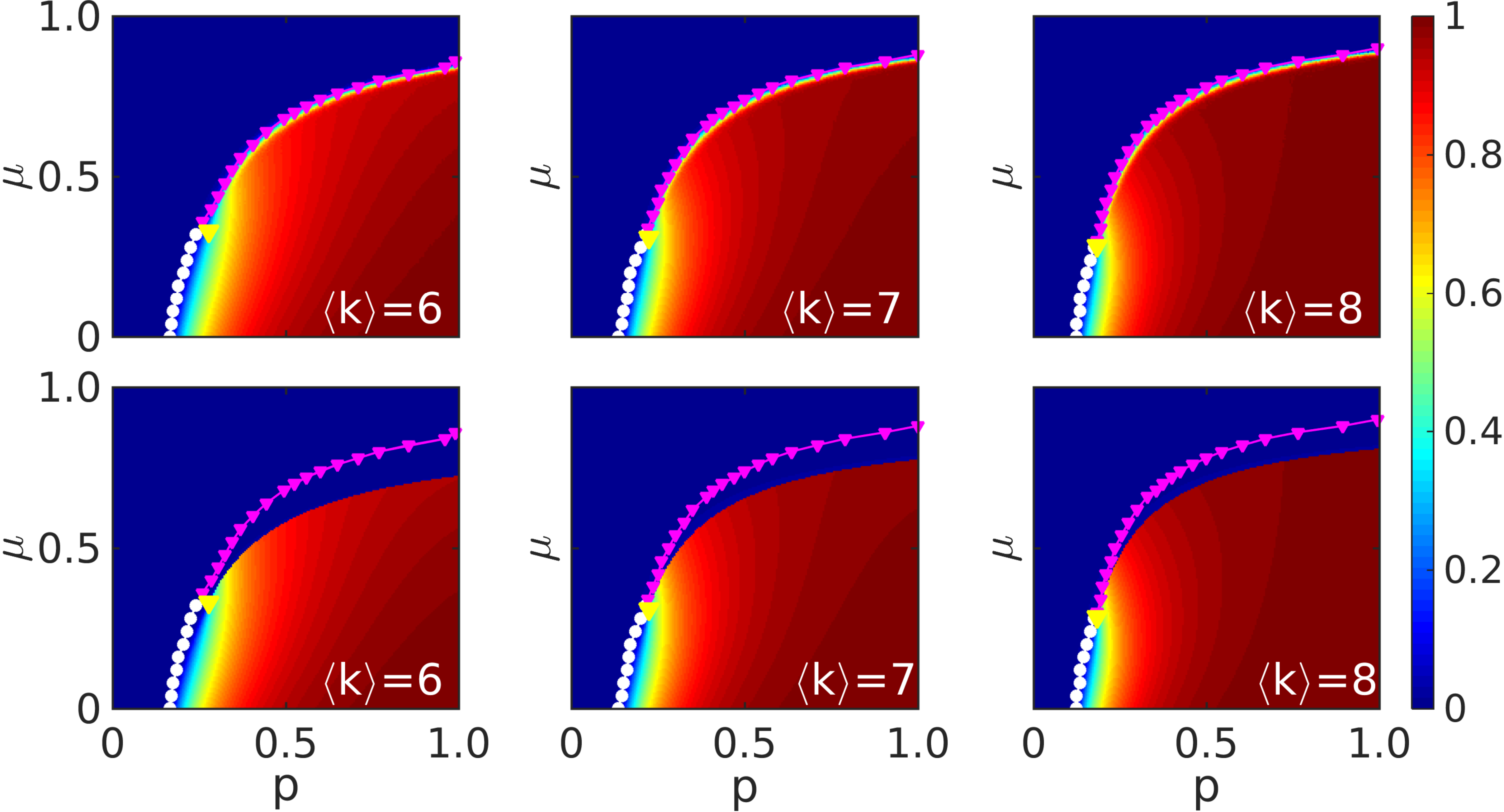}
\caption{ (Color online)
Effect of average degree on the contagion dynamics. Shown are the
color coded, numerically calculated (top panels) and theoretical predicted 
(bottom panels) final stationary distributions of recovered population for 
ER random networks in the parameter plane $(p,\mu)$. The values of the average 
degree tested are $\langle k\rangle=6$ (left column), $\langle k\rangle=7$ 
(middle column) and $\langle k\rangle=8$ (right column), with the corresponding
values of $p_e$ being $p_e=0.0006$, $p_e=0.0007$ and $p_e=0.0008$, respectively.
The main feature is that a relatively high average degree tends to facilitate 
contagion spreading. Other parameters are the same as those in 
Fig.~\ref{fig:simulation_all}.}
\label{fig:averagek}
\end{figure*}

Fig.~\ref{fig:averagek} shows, for ER random networks, the final stationary 
distributions of the recovered population for three values of the average 
degree $\langle k\rangle$ (or equivalently $p_e$). We observe that a relatively
high average degree tends to facilitate the expansion of the contagion. 
Especially, a large average degree means that the adopted individuals 
will have more chances to transmit the contagion to their connected susceptible neighbors, resulting in a larger contagion size and a smaller outbreak 
threshold. At the same time, for a fixed value of $\mu$, the individuals will 
have more close contacts, so that more susceptible individuals can reach the 
critical state, leading to relatively more massive adoption within a short 
time and an avalanche-like outbreak through subsequent transmission. Thus,     
as the value of the average degree is increased, more discontinuous
transitions occur at the critical boundaries. 

\subsection{Contagion dynamics on empirical networks} 
\label{subsec:realnetwork}

\begin{figure*}[!ht]
\centering
\includegraphics[width=\linewidth]{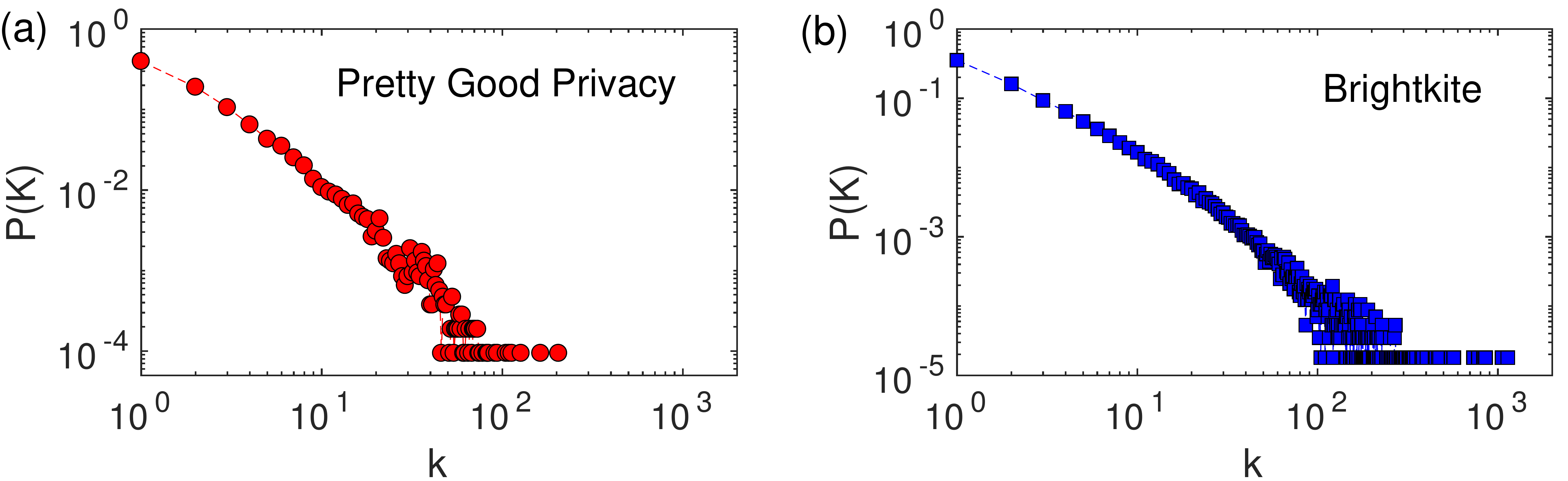}
\caption{ (Color online) Degree distributions of the two empirical 
networks. (a) The network Pretty Good Privacy (PGD). The structural parameters 
are $N=10680$, $\langle k\rangle=4.554$, $\langle k^{2}\rangle=85.976$,
maximum degree $k_{max}=205$, degree-degree correlation $r=0.238$, and
clustering coefficient $c=0.378$. (b) The network Brightkite. The structural
parameters are $N=56739$, $\langle k\rangle=7.506$,
$\langle k^{2}\rangle=480.61$, maximum degree $k_{max}=1134$, degree-degree 
correlation $r=0.01$, and clustering coefficient $c=0.111$.}
\label{fig:realdegree}
\end{figure*}

We test our spreading model using two empirical networks: an online contact 
network (Pretty Good Privacy (PGD)~\cite{pgd}) and the social network 
Brightkite~\cite{brightkite}. Both networks have a heterogeneous degree 
distribution with a substantial amount of clustering, as shown in 
Fig.~\ref{fig:realdegree}. In the simulations, the contagion starts 
from an adopted cluster of size $G(0)=30$.

\begin{figure*}[!ht]
\centering
\includegraphics[width=\linewidth]{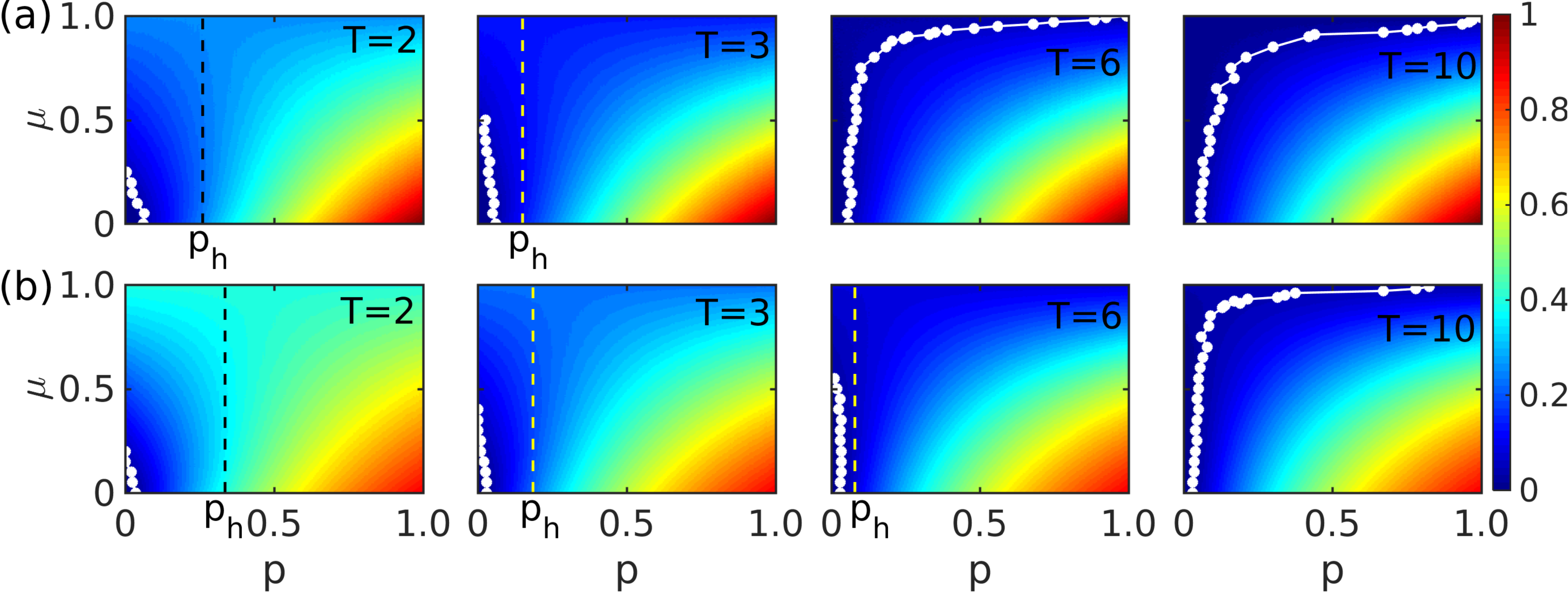}
\caption{ (Color online) Contagion dynamics and phase transition 
on the empirical network PGD (Pretty Good Privacy). (a) Numerically obtained 
final stationary distributions of the recovered population. (b) The 
corresponding result from randomly-reconnected networks with the degree of 
each node unchanged (first-order null network model). In detail, a new first-order 
null network with the same degree sequence of PGD is built after every $25$ independent 
realizations of the spreading dynamics. The solid white circles 
indicate the numerically estimated phase boundaries with respect to the 
relative variance $\nu(\infty)$. The dashed lines indicate the position of 
$p_h$, which are absent for $T>6$ in the top panels and $T=10$ in the bottom 
panels because closed contacts tend to block the contagion when the value 
of $T$ becomes large.}
\label{fig:pgdsimulation} 
\end{figure*}

\begin{figure*}[!ht]
\centering
\includegraphics[width=\linewidth]{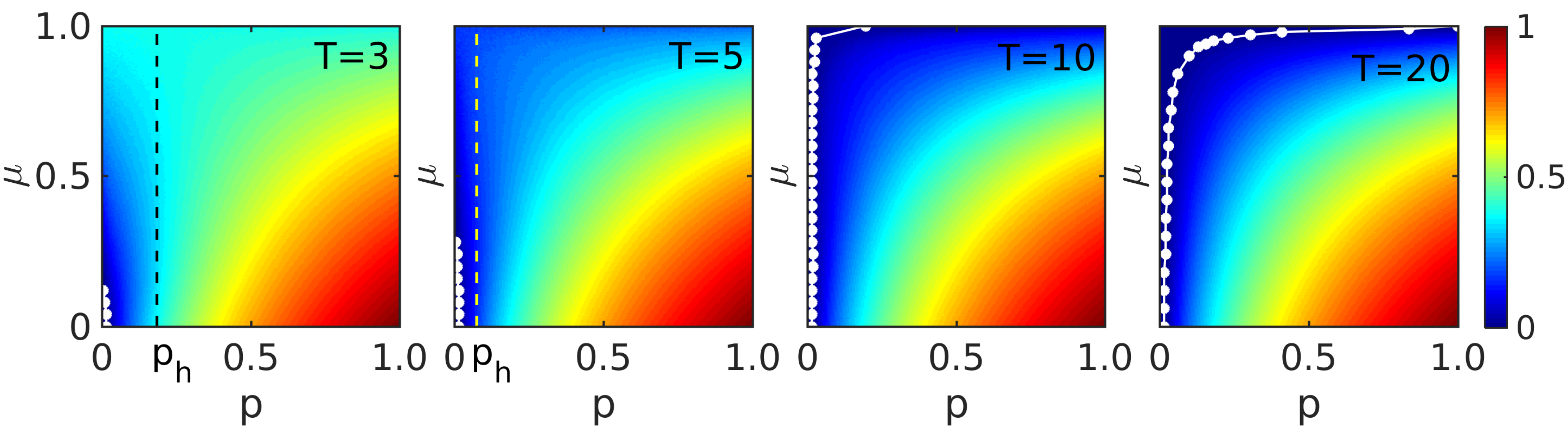}
\caption{(Color online) Contagion dynamics and phase transition 
on the empirical network Brightkite. Legends are the same as in 
Fig.~\ref{fig:pgdsimulation}. Due to the weak degree correlation ($r=0.01$) 
of this network, testing the first-order null network model is not essential.}
\label{fig:brightsimulation}
\end{figure*}

Study of the empirical networks has led to phenomena that are not present
in ER random networks, indicating the role of the network structure in the
social contagion dynamics. Representative results are shown in 
Figs.~\ref{fig:pgdsimulation} and \ref{fig:brightsimulation}.
Because of the heterogeneous connections of the individuals in the network, 
the phase transition is continuous, regardless of the value of $\mu$, in 
agreement with the findings of the previous works on threshold based dynamics 
in heterogeneous networks~\cite{Zhu2017,Wang2015,Chen2015}. In particular,
in such cases, the number of susceptible individuals' being able to reach 
a critical value within a short time is not sufficient to lead to an 
avalanche-like outbreak. Another phenomenon is that, when the fraction of 
closed contacts is large and the value of $T$ is relatively small, the 
threshold value is close to zero, suggesting that the outbreak of the 
contagion is largely driven by the hubs in the networks. However, the contagion size 
near the outbreak threshold is finite.

A heuristic mechanism leading to the phenomena in 
Figs.~\ref{fig:pgdsimulation} and \ref{fig:brightsimulation} can be described
as follows. For a small value of $T$, there can be two different cases in 
terms of two parameter regions divided by a numerically estimated threshold 
$p_h$, where $p_h$ indicates the position at which the quantity $\Delta(p)$ 
begins to change from being positive to being negative. The results in 
Fig.~\ref{fig:pgdsimulation} imply the following formula for determining the 
value of $\Delta(p)$: 
\begin{eqnarray} \label{eq:sumdelta}
\Delta(p)=\sum\limits_{\mu}(R(p,\mu+\delta)-R(p, \mu));
\end{eqnarray}
where $\Delta(p)>0$ [$\Delta(p)<0$] means that the final contagion size 
increases (decreases) gradually with $\mu$. This argument relies on one 
condition: for a fixed value of $p$, the final fraction $R(p,\mu)$ of the 
recovered population should change monotonically with $\mu$, which has been 
verified numerically.

\begin{figure*}[!ht]
\centering
\includegraphics[width=0.9\linewidth]{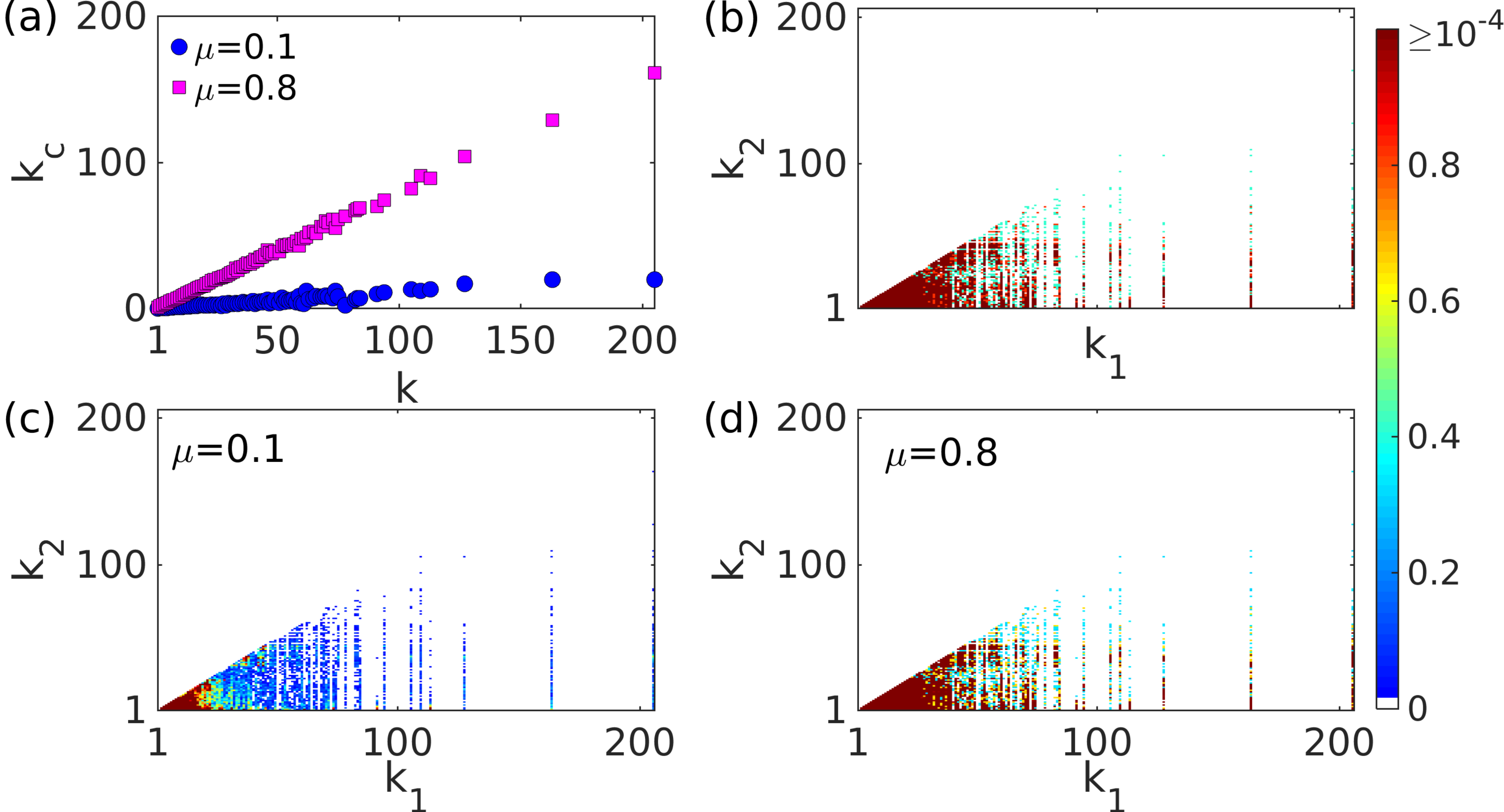}
\caption{ (Color online) 
Statistical behaviors of close contacts in the PGD 
network. Three key statistical characterizing quantities of close contacts 
are shown: (a) the expected number of close contacts that each node in 
different degree class can own for two different values of $\mu$, (b) the 
distribution of the edges between two nodes (node $1$ and node $2$, where 
$k_1$ and $k_2$ are the degrees of the two end nodes of an edge), normalized by 
the total number $E$ of edges in the network, and (c) the distribution of 
the close contacts between two nodes for $\mu=0.1$, normalized by $E$. 
(d) The distribution of the close contacts between two nodes for $\mu=0.8$. 
Because of the symmetry in the distribution, only half of the matrix is 
given in (b-c), where each color point indicates the value of $d(k_1,k_2)$ 
or $d(k_2,k_1)$, and $d(k_1,k_2)$ is the fraction of edges with two nodes 
of degree $k_1$ and $k_2$, respectively. The pink squares in (a) indicate 
that those nodes of high degree have considerable numbers of close contacts 
- a great advantage for transmission. The contagion spreads readily among 
these hubs due to the sufficient close contacts among them. The adopted 
areas are localized for small values of $p$. Similar phenomena are also found for 
the first-order null network model of PGD.}
\label{fig:pgdlocal}
\end{figure*}

For $p<p_h$, abundant close contacts in the network can promote the contagion
and lead to an outbreak, in contrast to ER random networks where this happens 
only for $T=1$. Because of the random allocation strategy of closed contacts 
among all connections, hubs have a natural advantage to possess a considerable 
number of close contacts. Evidence for this is presented in 
Figs.~\ref{fig:pgdlocal}(a), which show 
the expected numbers of the close contacts of nodes in different degree classes 
with the probability of an edge to be a close contact being $\mu$. As a result,
the hubs can readily enter the adopted state. For a large value of $\mu$, 
there is a higher probability for the contagion to be successfully transmitted 
among the hubs through close contacts, especially when the ordinary contacts 
have not been engaged in the transmission process (for small values of $p$).
Evidence for this is presented in Figs.~\ref{fig:pgdlocal}(b-d). 
The hubs can then facilitate the transmission,
especially for networks with weak local clustering. Note that the closed 
contacts capable of successfully transmitting the contagion tend to locate 
near the hubs, especially for the empirical network with a high positive 
degree-degree correlation [Fig.~\ref{fig:pgdsimulation}(a)]. However, this 
localization effect is weakened in the first-order null network model of PGD, 
as the hubs have more chances to connect with non-hub nodes, making a larger 
adopted population possible, as shown in Fig.~\ref{fig:pgdsimulation}(b). 
Nonetheless, global contagion does not arise near the outbreak threshold 
because of the small value of $p$ - in this case contagion pathways of long 
distances from the hubs are less likely.

As the value of $p$ is increased toward $p_h$, more and more ordinary contacts
begin to engage in the transmission process. The contagion is again no longer
restricted to local regions, and global contagion can occur through the
transmission along the long-range ordinary contacts. By comparing with
the results for ER networks, similar dynamics of the contagion but without 
discontinuous transitions for large $T$ can be observed. In the same way, 
sufficiently many close contacts in the network can inhibit global contagion. 
In such cases, in a network without close contacts, contagion can be maximized insofar 
as the value of the transmission rate $p$ is sufficiently large.

The similar dynamical behaviors in the PGD network and its first-order
network model suggest that the contagion spreading depends more on the
network heterogeneity than the degree-degree correlation. In fact, the
correlation serves to delay the outbreak of the contagion, as evidenced
by the larger outbreak thresholds in Fig.~\ref{fig:pgdsimulation}(a) and
\ref{fig:brightsimulation}.

Two other phenomena in Figs.~\ref{fig:pgdsimulation} and ~\ref{fig:brightsimulation} are
(1) the threshold $p_h$ disappears for large values of $T$ and (2) the 
parameter regions of global prevalence tend to shrink somewhat with $T$. 
The reason for the former is that a larger threshold value for successful 
adoption makes it harder for not only ordinary but also hub nodes to adopt 
the contagion through close contacts. Especially, hub nodes no longer have 
the advantage to take lead in promoting the spreading. Outbreak of the 
contagion is thus delayed or even completely suppressed by abundant close 
contacts in the network, as for ER networks. Instead, ordinary contacts play 
an increasingly pivotal role in dominating the contagion prevalence.

\begin{figure*}[!ht]
\centering
\includegraphics[width=\linewidth]{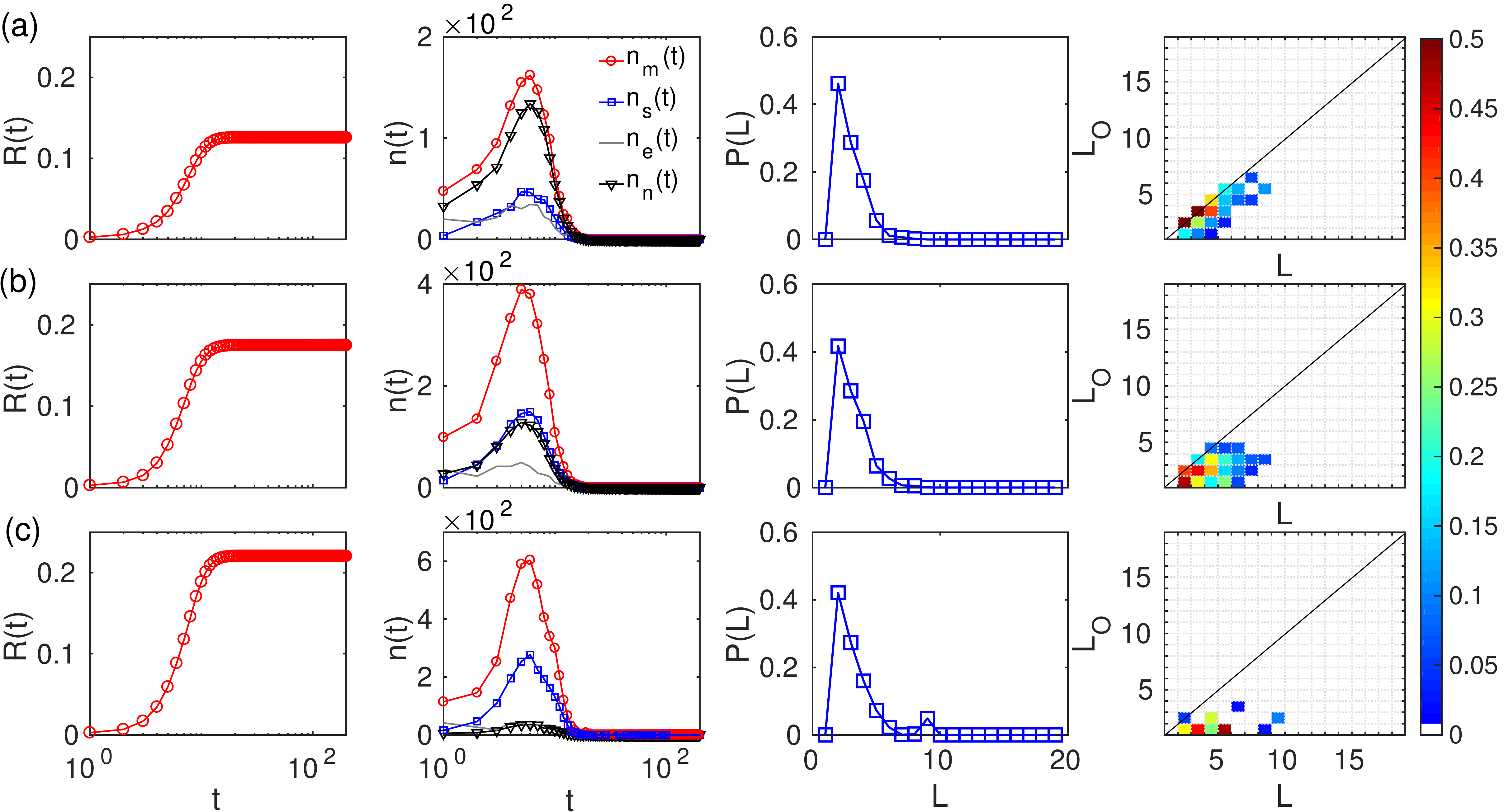}
\caption{ (Color online) 
The roles of two types of contacts in spreading dynamics 
in the PGD network. Shown are the behaviors of four statistical characterizing 
quantities for different values of $\mu$: (a) $\mu=0.1$, (b) $\mu_2=0.4$ and
(c) $\mu=0.8$, the positions of which are also labeled in 
Fig.~\ref{fig:transmission1}. Other parameter values are $p_h>p=0.16$, 
$q=1.0$ and $T=2$.}
\label{fig:pgdtransmission16}
\end{figure*}

\begin{figure*}[!ht]
\centering
\includegraphics[width=\linewidth]{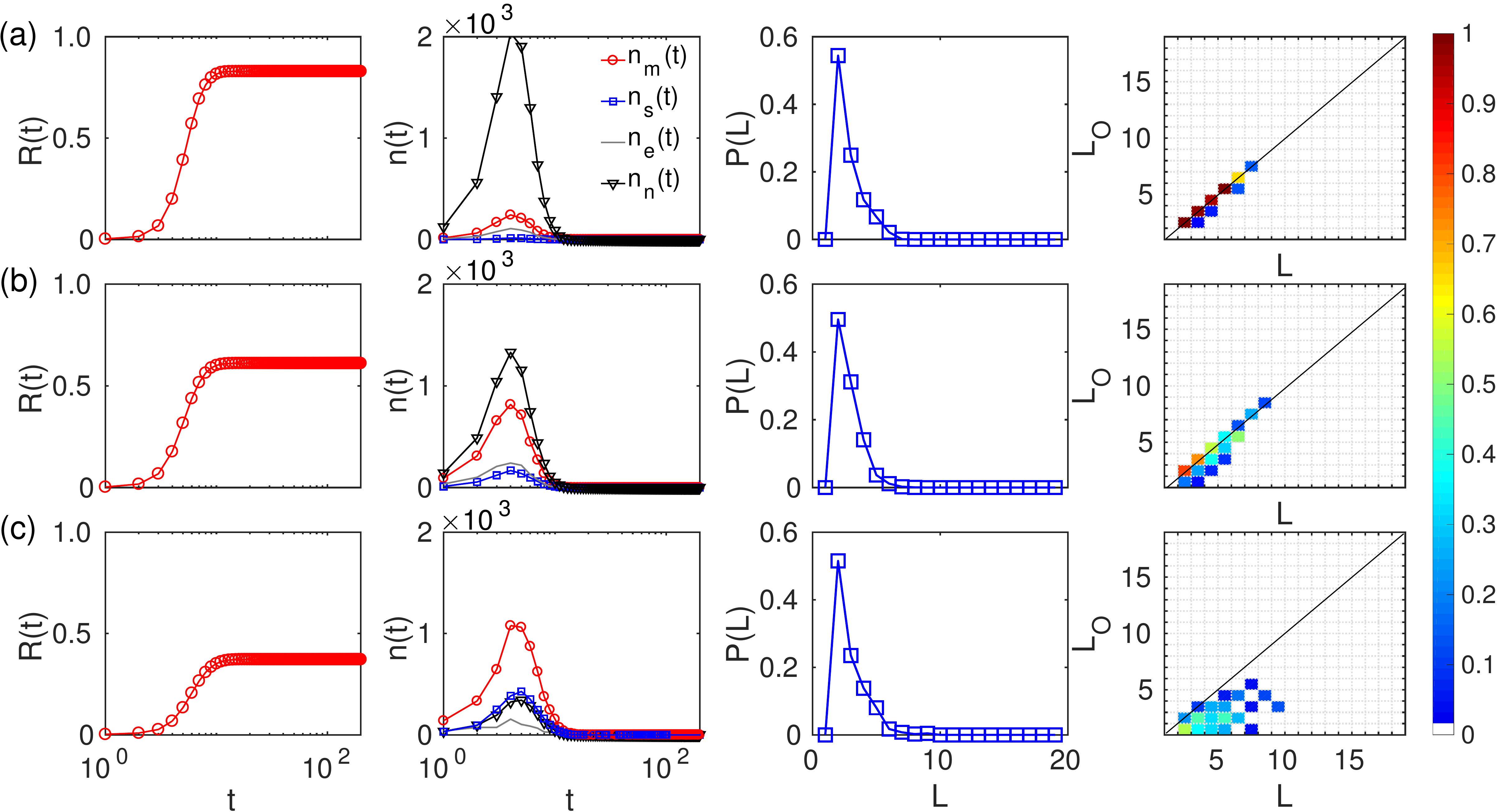}
\caption{(Color online) 
The roles of two types of contacts on spreading dynamics 
in the PGD network - additional support. Legends are the same as in 
Fig.~\ref{fig:pgdtransmission16} except for $p_h<p=0.9$.}
\label{fig:pgdtransmission9}
\end{figure*}

Surprisingly, the phenomenon that contagion is maximized when the roles of 
the two types of ties are comparable ceases to exist in the two empirical
networks. Some indications are shown in
Figs.~\ref{fig:pgdtransmission16} and \ref{fig:pgdtransmission9}, where the
third columns of the both figures suggest the absence of diffusion paths of
long length ($L>10$) and the dominance of short paths. This is due to the lack
of long-time cooperation between successful reinforced and ordinary
transmission, as shown in the forth columns of
Figs.~\ref{fig:pgdtransmission16} and \ref{fig:pgdtransmission9}. In
particular, there is still cooperation between the two types of transmission
events, but it is limited to short diffusion paths (see the third columns of
Figs.~\ref{fig:pgdtransmission16} and \ref{fig:pgdtransmission9}), regardless of the value
of the transmission rate $p$. As a result, an avalanche type of spreading
process fails to persist at large scales. Overall, heterogeneous networks
do not provide the structural condition for cooperation between closed and
ordinary contacts along long paths.

\section{Discussion} \label{sec:discussion}

To summarize, empirical evidence and understanding of social contagion 
suggest the indispensable roles played by ordinary and close contacts in 
the spreading dynamics, calling for a general model to take into account 
both types of social contacts. We have accomplished that in this paper.   
In addition, we have developed a theoretical approach to analyzing the 
spreading dynamics of social contagion on random networks. 
Agent-based simulations and theoretical analyses have revealed the 
coexistence of both continuous and discontinuous phase transitions in 
ER random networks. Study of two empirical networks with a heterogeneous
degree distribution and a substantial amount of clustering has indicated  
that only continuous transitions can arise in such networks, and we have 
provided a physical understanding.
Our findings are consistent with previous results from threshold models 
incorporating some kind of social reinforcement 
mechanism~\cite{Wang2015,Zhu2017,Chen2015}. The physical origin of the 
discontinuous transition in a social system with a large number of close 
contacts can be attributed to their abundance through a drastic avalanche 
process~\cite{Wang2015,Cui2018}. 
A general result is that the value of the outbreak threshold 
is mainly determined by the abundance of the close contacts. 

For ER networks, our study has revealed that close social contacts alone 
are not capable of triggering a large scale outbreak of contagion. To have 
a global outbreak, a sufficient amount of ordinary transmission is needed. 
We have obtained a physical understanding of this phenomenon through a 
detailed statistical analysis of the spreading dynamics in terms of the 
frequencies of three different types of transmission modes, the distribution 
of diffusion paths of various lengths, and the frequencies of the occurrence 
of transmission along the diffusion paths. The analysis has revealed that, 
to achieve maximum spreading, a matching condition must be met: 
$n_{O}(t)\approx n_{D}(t)$, i.e., the number of close contacts available to 
susceptible individuals in the critical state must approximately match the 
number of ordinary contacts. There exists an optimal fraction of close 
contacts to maximize the spreading, which depends on the values of the 
transmission rate and adoption threshold. The key to making spreading 
prevalent lies in the interplay between ordinary and successfully reinforced 
transmission events associated with long diffusion paths, which cannot occur 
unless there is a sufficient amount of ordinary transmission. If ordinary 
contacts are scarce, spreading will become stagnant. Ordinary contacts thus 
play a crucial role in promoting social contagion when both types of social 
contacts are simultaneously present. In addition, a sufficient number of 
close contacts can change the nature of the phase transition from continuous 
to discontinuous, and leads to an increase in the value of the outbreak 
threshold.

For heterogeneous networks, a different picture of contagion spreading dynamics 
arises. While ordinary contacts still play a dominating role in the global 
prevalence of the contagion, there is a sensitivity to the threshold required 
for a successful adoption along close contacts. In particular, for a small 
threshold value, abundant close contacts facilitate the outbreak due to the
availability of the transmission channels among the hub nodes, regardless of 
the value of the transmission rate. Close contacts are capable of promoting 
spreading but only on a local scale, especially near the hubs, making it 
difficult to achieve global prevalence of the contagion even though contagion 
outbreaks can still occur. For a large threshold value, close contacts tend 
to become an obstacle to transmission: global contagion will be prevented 
if the network is rich in close contacts, which is similar to what happens 
in ER networks. The lack of avalanche-like spreading, which results from 
inadequate number of susceptible individuals in the critical state, imposes 
a limit to the cooperation between ordinary and successful reinforced 
transmission induced by two types of contacts along long paths, leading to 
the absence of an optimal intermediate value $\mu$ to maximize the contagion 
spreading.

We conclude our work by providing two general remarks. Firstly, numerical 
computations are inadequate to conclude the coexistence of continuous and 
discontinuous transitions in ER networks. Nonetheless, the issue for ER 
networks can be settled analytically through a comprehensive bifurcation 
analysis of the system dynamics using theoretical methods such as the 
edge-based compartmental approach~\cite{Parshani2010}, which we have carried
out. Secondly, the main goal of our present study is to develop a computational
and theoretical paradigm to understand the relative roles played by ordinary 
and close contacts in the social contagion. For this reason, 
we have studied the process involving a single contagion for both static 
random and empirical heterogeneous networks. 
We hope to be able to extend our analysis of either coupled spreading 
scenarios~\cite{Claudio2017} or contagion dynamics to more diverse complex 
networks, as well as to temporal networks~\cite{Holme2012}, multilayer 
networks~\cite{Kivela2014,Boccaletti2014,Battiston2017,Chen20181,
LWCTL:2018}, and metapopulation systems~\cite{Wang2018}. 

\section*{Acknowledgments}

This work was partially supported by China Postdoctoral Science Foundation 
(PSF) under Grants No.~2015M582532 and No.~2018M631073, by Fundamental Research 
Funds for the Central Universities (No.~YJ201830), by Science Strength 
Promotion Programme of UESTC and by the National Natural Science Foundation 
of China under Grants No.~61673086 and No.~61433014. 
YCL would like to acknowledge
support from the Vannevar Bush Faculty Fellowship program sponsored by the
Basic Research Office of the Assistant Secretary of Defense for Research and
Engineering and funded by the Office of Naval Research through Grant
No.~N00014-16-1-2828.


%

\newpage

\end{document}